\documentclass[]{aa} 
\usepackage{graphicx}
\usepackage{natbib}
\bibpunct{(}{)}{;}{a}{}{,} 
\newcommand{\TO}[0]{\mbox{$-$}}
\newcommand{\MOD}{}   
\newcommand{\NEW}{}   
\newcommand{\RM}[1]{\mathrm{#1}}
\newcommand{\OUT}[1]{{\tiny }}
\newcommand{\LIT}[1]{}
\newcommand{\HIDE}[1]{}
\newcommand{\NOTE}[1]{^{\mathrm{{\it #1}}}}
\newcommand{\kkms}[0]{\RM{K}\,\RM{km}\,\RM{s}^{-1}}
\newcommand{\kms}[0]{\,\RM{km}\,\RM{s}^{-1}}
\newcommand{\micron}{\mbox{$\mu$m}}

\def\percc {$\hbox{{\rm cm}}^{-3}$}
\newcommand{\CIIw}{\ion{C}{ii}}

\newcommand{\CII}{[\ion{C}{ii}]}
\newcommand{\CIw}{\ion{C}{i}}
\newcommand{\CI}{[\ion{C}{i}]}
\newcommand{\OIw}{\ion{O}{i}}
\newcommand{\OI}{[\ion{O}{i}]}
\newcommand{\OIII}{[\ion{O}{iii}]}

\newcommand{\HII}{\ion{H}{ii}}

\newcommand{\NII}{[\ion{N}{ii}]}
\newcommand{\msol}{M$_\odot$}

\newcommand{\cmsq}{cm$^{-2}$}

\defcitealias{schneider2006}{SBS2006}

\begin{document}

\title{The cooling of atomic and molecular gas in DR21}
   \titlerunning{The cooling of atomic and molecular gas in DR21}
  
   \author{H.\,Jakob\inst{1} \and 
          C. Kramer\inst{1} \and
          R. Simon\inst{1} \and
          N. Schneider\inst{4,2,1} \and
          V. Ossenkopf\inst{1,3} \and
          S. Bontemps\inst{2} \and
          U.~U. Graf\inst{1} \and
          J. Stutzki\inst{1}
          }
   \institute{KOSMA, I. Physikalisches Institut,
              Universit\"at zu K\"oln,
              Z\"ulpicher Stra\ss{}e 77,
              50937 K\"oln, Germany
              \and
              Observatoire de Bordeaux,
              Universit\'e de Bordeaux 1,
              BP 89, 33270 Floirac, France
              \and
              SRON, National Institute for Space Research,
              PO Box 800, 9700 AV Groningen, the Netherlands
              \and
              SAp/CEA Saclay, 91191 Gif-sur-Yvette, France
             }
   \offprints{H.\,Jakob}
   \mail{jakob@ph1.uni-koeln.de}
   \date{Received 19 June 2006 / Accepted 11 September 2006}

   \abstract
     {}
     { We present an overview of a high-mass star formation region through the major (sub-)mm, and far-infrared
       cooling lines to gain insight into the physical conditions and the energy budget of the molecular cloud.
     }
     {
     We used the KOSMA 3m telescope to map the core
     ($10'\times 14'$) of the Galactic star forming region DR\,21/DR\,21\,(OH) in the Cygnus X region
     in the two fine structure lines of atomic carbon (\CIw{} $^3\RM{P}_1\TO^3\RM{P}_0$ and $^3\RM{P}_2\TO^3\RM{P}_1$) and four mid-$J$
     transitions of CO and $^{13}$CO, and CS $J=7\TO6$.
     These observations have been combined with FCRAO $J=1\TO0$ observations of $^{13}$CO and C$^{18}$O.
     Five positions, including DR21, DR21\,(OH), and DR21\,FIR1, were observed 
     with the ISO/LWS grating spectrometer in the \OI\ 63 and 145 $\mu$m lines, the \CII\ 
     158 $\mu$m line, and four high-$J$ CO lines. We discuss the intensities 
     and line ratios at these positions and apply Local Thermal Equilibrium (LTE)
     and non-LTE analysis methods in order to
     derive physical parameters such as masses, densities and temperatures.
     The CO line emission has been modeled up to $J=20$.
     }
     {
     From non-LTE modeling of the low- to high-$J$ CO lines we identify two gas components, a cold one at temperatures of 
     T$_\RM{kin}\sim 30-40$\,K, and
     one with T$_\RM{kin}\sim 80-150$\,K at a local clump density of about n(H$_2$)$\sim 10^4-10^6$\,cm$^{-3}$.
     {{\NEW While the cold quiescent component is massive containing typically more than 94\,\% of the mass, the warm,
     dense, and turbulent gas is dominated by mid- and high-$J$ CO line emission and its large line widths.
     }}
     The medium must be clumpy with a volume-filling of a few percent.
     The CO lines are found to be important for the cooling of the cold molecular gas, e.g. at DR21\,(OH).
     Near the outflow of the UV-heated source DR21, the gas cooling is dominated by line emission of atomic oxygen
     and of CO.
     Atomic and ionised carbon play a minor role.
     }
     {}

  \keywords{ISM: clouds -- ISM: abundances -- Radio lines: ISM -- Line: profiles -- Stars: formation -- ISM: individual objects: DR\,21, DR21\,(OH)}
   \maketitle
%
%
%
\begin{figure*}[ht]
\begin{center}
\includegraphics[width=9.05cm,angle=-90]{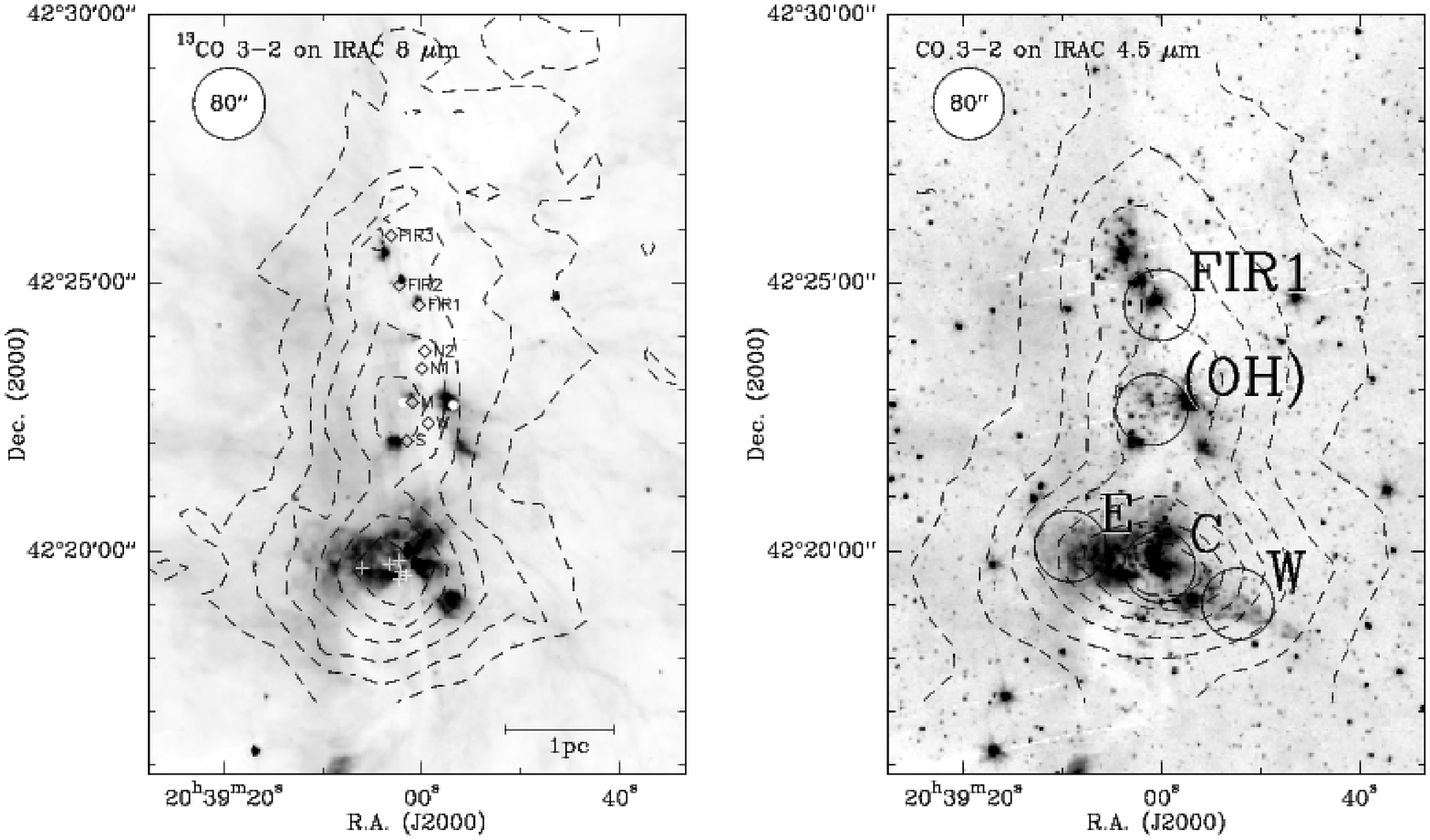}
\end{center}
\caption{
View of the DR21/DR21\,(OH) high mass star forming region.
{\it Left:} Overlay of the KOSMA $^{13}$CO 3\TO2 map (cf. \citet{schneider2006}/\citetalias{schneider2006}) in contours 
on a Spitzer IRAC 8\,\micron{}
greyscale map \citep{marston2004}.
Contours are on a T$_\RM{mb}$ scale and range from
16.8 to 100.8 in steps of 16.8\,$\kkms$ ($\sigma=5.6\,\kkms$).
The integration interval is
$-10$ to $20\,\kms$
The left-top circle indicates the HPBW of KOSMA.
The map also shows 6 positions of radio continuum sources identified as OB stars by \citet[][]{roelfsema1989} (white crosses),
2 infrared sources and maser positions \citep[][]{garden1986} (small white circles),
and 8 compact 1.3-mm dust emission sources \citep{chandler1993} (open rectangles).
{\it Right:} Overlay of the Spitzer map at $4.5$\,\micron{} with contours of the KOSMA $^{12}$CO 3\TO2 map (integration interval of
$-20$ to $30\,\kms$; discussed in \citetalias{schneider2006}) ranging from
18 to 450 in steps of 48\,$\kkms$ ($\sigma=6\,\kkms$).
The labeled circles mark the positions and beam sizes of the ISO/LWS observations that we use to constrain the physical paramters as discussed in the text.
The LWS beam is comparable to the resolution of the KOSMA maps.
\label{fig_spitzer}
}
\end{figure*}

%

\begin{table*}[ht]
\caption[]{List of observational parameters for KOSMA.
The columns give line frequency, {\NEW receiver used}, main beam efficiency $\eta_{\RM{mb}}$,
half power beam width, observing mode, number of mapped positions, 
velocity channel width $\Delta$v (the resolution is a factor of
$\sim1.4$ larger), mean atmospheric zenith opacity 
$\langle\tau^\mathrm{atm}_0\rangle$, and observing period.
}
\label{tab-obslist}
\begin{tabular}{llcccclrccr}
\noalign{\smallskip} \hline \hline \noalign{\smallskip}
Species & Transition & Frequency &Receiver& $\eta_\RM{mb}\NOTE{a}$ & HPBW$\NOTE{a}$ &
  Mode & No. & $\Delta$v & $\langle\tau^\RM{atm}_0\rangle$ &
  Observing period \\
  &&[GHz]&&&[$''$]&&&[m s$^{-1}$] & & \\
\noalign{\smallskip} \hline \noalign{\smallskip}
$^{13}$CO$\NOTE{b}$  & $J=3\TO2$           & 330.5880 &DualSIS& 0.68 & 80 & OTF$\NOTE{c}$
    & 693 & 308 & 0.26 & 01/2006\\
$^{12}$C$^{32}$S  & $J=7\TO6$           & 342.8829 && 0.68 & 80 & OTF$\NOTE{c}$
    & 525 & 294 & 0.08 & 12/2005 \\
$^{12}$CO  & $J=3\TO2$           & 345.7960 && 0.68 & 80 & OTF$\NOTE{c}$
    & 525 & 294 & 0.08 & 12/2005 \\
$^{12}$CO  & $J=4\TO3$           & 461.0408 &SMART& 0.5 & 57 & OTF$\NOTE{c}$
    & 3458 & 677 & $0.85-1.4$ & 1/2005\\
$[^{12}$\CIw] & $^3$P$_1\TO^3$P$_0$ & 492.1607 && 0.5 & 55 & DBS$\NOTE{d}$
    &  1380 & 630 & $0.5-1.6$ & 1/2003 -- 3/2004\\
$^{13}$CO  & $J=6\TO5$           & 661.067 && 0.4 & $\sim$$40$ & PSw$\NOTE{e}$
    & 33 & 307 & $1.0-1.3$ & 1/1998 \\
$^{12}$CO  & $J=6\TO5$           & 691.473 && 0.4 & $\sim$$40$ & PSw$\NOTE{e}$
    & 58 & 295 & $1.0-1.3$ & 1/1998 \\
$^{12}$CO  & $J=7\TO6$           & 806.6518 && 0.31 & 42 & DBS$\NOTE{d}$
    & 964 &  775 &  $0.4-1.6$ & 1/2003 -- 3/2004 \\
$[^{12}$\CIw] & $^3$P$_2\TO^3$P$_1$ & 809.3420 && 0.31 & 42 & DBS$\NOTE{d}$
    &  964 & 775 & $0.4-1.6$ & 1/2003 -- 3/2004\\
\noalign{\smallskip} \hline \end{tabular}
\begin{list}{}{}
\item[$\NOTE{a}$] The main beam efficiency $\eta_\RM{mb}$ and the HPBW were
    determined from cross scans of Jupiter.
\item[$\NOTE{b}$] See \citet{schneider2006} for a $^{13}$CO 3\TO2 map covering a larger part of Cygnus X.
\item[$\NOTE{c}$]  OTF -- On-The-Fly mapping mode with periodic switching to a emission free reference position at\\
$\alpha(\RM{J2000})=20^\RM{h}37^\RM{m}10^\RM{s}$ $\delta(\RM{J2000})=42^\circ30'00''$.
\item[$\NOTE{d}$]  DBS -- Dual Beam Switch observing mode with a chopping secondary mirrow in azimuth.
\item[$\NOTE{e}$]
The CO and $^{13}$CO $6\TO5$ observations were
 carried out on a $35''$
grid \citep{koester1998} in a  Position-Switched\\
observing mode (PSw).
The OFF position was at
$\alpha(J2000)$=$20^\RM{h}26^\RM{m}43^\RM{s}.56$,
  $\delta(J2000)$=$42^\circ09'39''.5$.
\end{list}
\end{table*}

%

\section{Introduction}

Emission lines of \CIIw{}, \CIw{}, and CO in molecular clouds are among the most important diagnostic probes of star formation and particularly important
for studying the influence of massive stars and their UV radiation on the environment.
Obtaining the physical parameters of such regions
is often done with simplified radiative transfer assumptions.
Local Thermal Equilibrium (LTE) and the Large Velocity Gradient (LVG) analysis have
proven to yield good results for the bulk of the cold gas (see, e.g., \citet{black2000}).
However, the assumption of a homogeneous medium does not represent the typical physical conditions of a high mass star formation region.
{\NEW
Instead,
the observing beam will contain several gas components with different
excitation conditions and with unknow filling factors.
The observed line transitions are sensitive to gas at different density, temperature and opacity,
corresponding to different regions.
}
%
{\NEW
By selecting designated
lines as tracers of
}
the different parts of the gas,
one can obtain a rather complete coverage of excitation conditions accounting for different physical and also chemical conditions
{\NEW within the beam}.
A more differenciated analysis 
using non-LTE radiative transfer algorithms (see \citet{vanzadelhoff2002} for an overview) is needed for this purpose. It can provide
a more detailed description of the structure of the cloud including clumpiness and temperature or density
gradients (e.g., \citet{ossenkopf2001}, \citet{williams2004}).
\OUT{
Although optically thin thermal dust emission at mid- to far-IR wavelengths is a valuable tool to determine the column density and
the temperature it may not resolve the velocity structure or give an estimate of the local density.
}
%

Using the results of this analysis, we can try to retrieve a self-consistent 
picture including all major processes driving and driven by high mass star
formation (e.g, with respect to turbulence or chemistry).
{\NEW Thus,} we obtain a detailed view of the Photon Dominated Regions (PDRs)
{\NEW in}
the environment of young stars. PDRs are transition regions from
ionized and atomic to dense molecular gas in which physics and chemistry
{\NEW of} the gas are dominated by the UV radiation from young massive stars.
Processes heating and cooling the clouds are qualitatively well understood
on a theoretical level \citep[e.g.,][]{tielens1985,sternberg1989},
and in general we find a reasonable agreement between the theoretical
predictions and observations of the brightest cooling lines.
Still, there is an ongoing debate about the details of the energy input
contributing to the heating. We find a combination of
mechanisms such as stellar far-ultraviolet radiation (FUV),
dynamical energy through shocks or supernovae, or
an enhanced cosmic ray rate. There are also many open questions
with respect to the chemical composition of these regions.
The role of nonequilibrium chemistry is still not well understood.
For instance Polycyclic Aromatic Hydrocarbons (PAHs) as a reservoir
of fresh gas phase carbon (e.g., \citet[][]{oka2004}, \citet{habart2005})
may change the overal picture.

{\NEW Emission from warm carbon in the form of atoms, ions or in carbonated molecules (mainly CO) provides a significant fraction of the gas cooling
at sub-millimeter and far-infrared (FIR) wavelengths}
(e.g.,
\citet{giannini2000,giannini2001},
\citet[][]{schneider2003},
\citet{kramer2004},
or \citet{bradford2003,bradford2005}).
Finding the prevailing cooling channels among these and other species
(e.g., \OI, H$_2$O, OH)
may provide significant insight into the details of the chemical
composition and the energy balance of PDRs.
Since systematic observations of sub-mm- and FIR-line cooling lines
with a good spatial resolution
{\NEW are still sparse, data to test the model predictions are still largely
missing and many questions concerning the details of the chemistry and energy
balance remain unanswered. By studying a well-known, massive Galactic star forming region,
we aim at gaining new insight into these problems.}

%
\OUT{
 the mid- and high-$J$ CO line emission was assigned to 
a high temperature gas component at densities of $10^5$\,\percc{} and higher.
This component constitutes only a few percent of the total cloud mass and
is thus not representative for the analysis of the (cooler) main gas component.

In order to address this problem we need to pin down the origin of the line emission.
CO rotational lines proved to be good tracers of the molecular gas.
This was shown many times in comparison studies to other surrogates as well as to dust emission studies \citep[e.g., ][]{goldsmith1997}.
Collisions with H$_2$ molecules and the radiative transfer through the local radiation
field dominate the CO excitation. They depend on the kinetic temperature and density structure of a cloud.
Hence, consistent modeling using radiative transfer methods should tell us about the physical conditions in the cloud.
}
%
%
%
\OUT{ 
Theoretical \citep{tielens1985,sternberg1989} and observational 
studies (references) all come to the same conclusion of a layered structure: the ionising photons 
first create a zone (at a visual extinction of A$_v\sim$1$^m$) where hydrogen and 
atomic oxygen are neutral and atomic carbon ionised. Deeper penetration into
in a clumpy molecular medium \citep[ and other references]{stutzki1988}
leads to an enriched zone of molecular gas where 
carbon first becomes neutral, and the increasing attenuation of the UV radiation leads to
a warm molecular zone of high excitation (at A$_v\sim$2--5$^m$), followed by the cloud core (A$_v\sim$10$^m$)
with carbon mostly contained in CO molecules.
The brightest cooling lines in the sub-millimeter and far-infrared (FIR) wavelength regime for the gas are 
the \CII\ 158\,\micron{}, \OI\ 63\,\micron{} finestructure lines, the CO rotational 
lines, H$_2$ near infrared lines, and numerous Polycyclic Aromatic Hydrocarbon (PAH) features.
They are crucial to determine the physical parameters of the PDR.

The common approach to study PDRs is to combine observed line intensities and ratios from 
PDR tracers with theoretical calculations in order to reproduce the observations by 
varying the hydrogen density and intensity of the UV radiation field. Several studies 
of high mass star forming regions have already been presented (references) but many lack 
a sufficient number of observed species and positions in order to pinpoint exactly the 
performance and limitations of existing PDR models. 
}

\OUT{
In this paper, we present maps and single pointings of a large number of PDR tracers 
in the DR21 region (see below) in order to examine the physical properties of this star forming region. 
Particularly multi-wavelength studies enable to draw a coherent picture of
the processes driving and driven by high mass star formation.
Thus, large-scale surveys in \CI\ and CO emission obtained with, e.g., the KOSMA, Mt.Fuji, and AST/RO telescopes
in conjunction with the ISO or Kuiper Airborne Observatory (and in future with Herschel) far-IR data 
provide a useful diagnostic of the conditions associated with active massive star formation
\citep[e.g.,][]{schneider2003,kramer2004,yamamoto2001,oka2004,zhang2001}.
}

\subsection*{The DR\,21 region}

The DR\,21 \HII-region/molecular cloud is part of the Cygnus~X complex of molecular clouds.
Cygnus~X is known to host a large number of \HII{}-regions 
\citep{downes1966,wendker1984} associated with molecular clouds
as seen, e.g., in the CfA $^{12}$CO 1\TO0 survey (\citet{leung1992}).
An extended (11 square degrees) map of the Cygnus X region in the
$^{13}$CO 2\TO1 line and smaller maps in the CO and $^{13}$CO 3\TO2 lines were obtained with the KOSMA telescope and are presented
in \citet{schneider2006} (from now on \citetalias{schneider2006}).
The distance to DR\,21 was sometimes estimated to be $\sim3$\,kpc,
based on visual interstellar extinction and radial velocity \citep[e.g.,][]{campbell1982}.
For this paper, we follow the argumentation from \citetalias{schneider2006} based on a morphological comparison of the molecular line
data with mid-IR emission and the OB associations, and favour a value of
1.7\,kpc as distance toward DR21 and DR21\,(OH). However, for easier comparison with the literature, we explicitly
give distance depended physical properties as a function of distance.

The DR\,21 region has been subject to numerous studies at different wavelengths.
Several active star formation sites and cometary shaped \HII{}-regions were identified
in maps of radio continuum emission \citep[e.g., ][]{dickel1983,cyganowski2003}.
Ammonia observations by \citet[][]{mauersberger1985} were interpreted
as evidence for high densities ($\sim$$10^{5.5}$\,\percc{}) paired with a high degree of clumpiness.
H$_2$O masers \citep{genzel1977} and a
recently obtained map of 1.3-mm continuum emission \citep{motte2005}
show that DR\,21 belongs to a north-south orientated
chain of {\sl massive star forming complexes}.
Additional indication of active star formation is also given by observations of vibrationally
excited H$_2$, tracing the hot, shocked gas in the vicinity of DR\,21 and DR21\,(OH)
(e.g., \citet{garden1986}, 
{\NEW and more recent results by \citet{davis2006}}).
H$_2$ images reveal an extended ($5'$=2.5\,pc(D/1.7\,kpc)) north-east
to south-west orientated outflow centered on DR21.
To the west, the outflow is highly collimated, whereas toward the east side, gas is expanding in a blister style \citep{lane1990}.
\citet{jaffe1989} found very intense CO 7\TO6 emission peaking on the \HII{}-region and following the H$_2$-emission.

%

In this paper, we present the results of sub-mm mapping 
observations with KOSMA and discuss far-IR line intensities observed with ISO.
This comprehensive data set allows to study the spatial and partly also the
kinematic structure of all major cooling lines of the ISM in the vicinity of DR\,21 and DR\,21(OH).
In Sect. 3, we analyse the line emission and continuum data toward five positions,
by comparing the spatial distribution, line profiles, and the physical properties 
through selected line ratios and the far-IR dust continuum.
In Sect. 4, we model the integrated line intensity ratios and absolute intensities
with radiative transfer models.
The impact of the line emission on the total cloud cooling is dicussed in Sect. 5.
A summary is given in Sect. 6.

%

%

\begin{table}[ht]
\caption[]
{Positions of ISO observations in absolute coordinates.
DR21\,C{} is centered on the \HII{}-region.
Projected distance of DR21\,E{} (resp. DR21\,W{}) from position DR21\,C{} is
0.77\,pc (resp. 0.89\,pc) at a distance of 1.7\,kpc.
The TDT-number is given in the second column.

\label{tab-iso-positions}}
\begin{tabular}{lclcc}
\noalign{\smallskip} \hline \hline \noalign{\smallskip}
Name & TDT-No. & R.A. [J2000] & Dec. [J2000]\\ 
\noalign{\smallskip} \hline \noalign{\smallskip}
DR\,21 E{}/East & 15200785 & $20^\RM{h}39^\RM{m}09^\RM{s}.14$ &$42^\circ20'05''.0$ \\
DR\,21 C{}/Center & 15200786 & $20^\RM{h}39^\RM{m}00^\RM{s}.93$ &$42^\circ19'42''.0$ \\
DR\,21 W{}/West & 15200787 & $20^\RM{h}38^\RM{m}52^\RM{s}.23$ &$42^\circ19'00''.5$\\
DR\,21 (OH)& 34700439 & $20^\RM{h}39^\RM{m}00^\RM{s}.95$ &$42^\circ22'37''.6$ \\
DR\,21 FIR1& 35500317 & $20^\RM{h}39^\RM{m}00^\RM{s}.2$  &$42^\circ24'34''.6$\\
\noalign{\smallskip} \hline \noalign{\smallskip}
\end{tabular}
\end{table}

%

\begin{figure*}[ht]
\includegraphics[width=9.05cm,angle=-90]{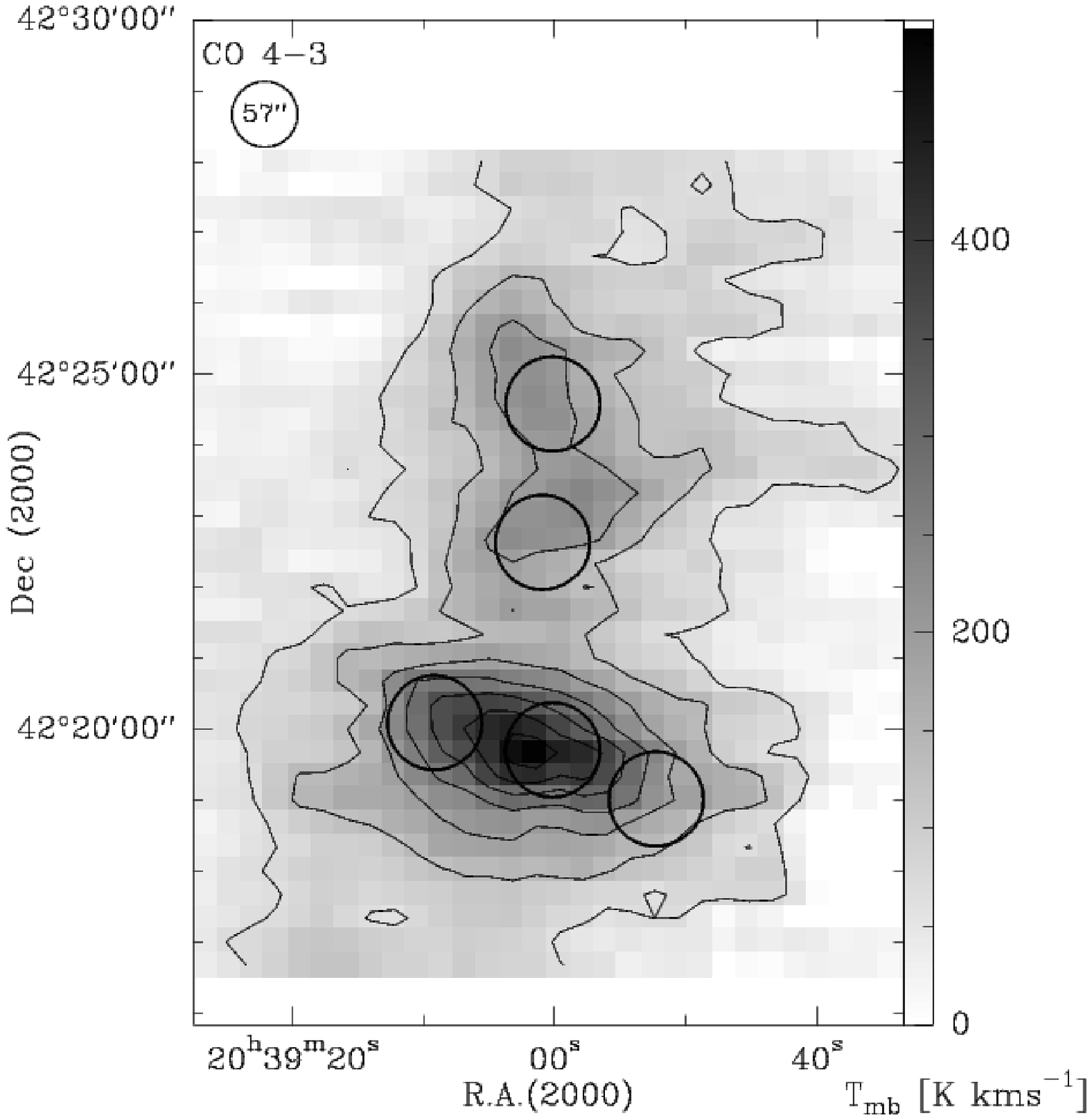}
\includegraphics[width=9.05cm,angle=-90]{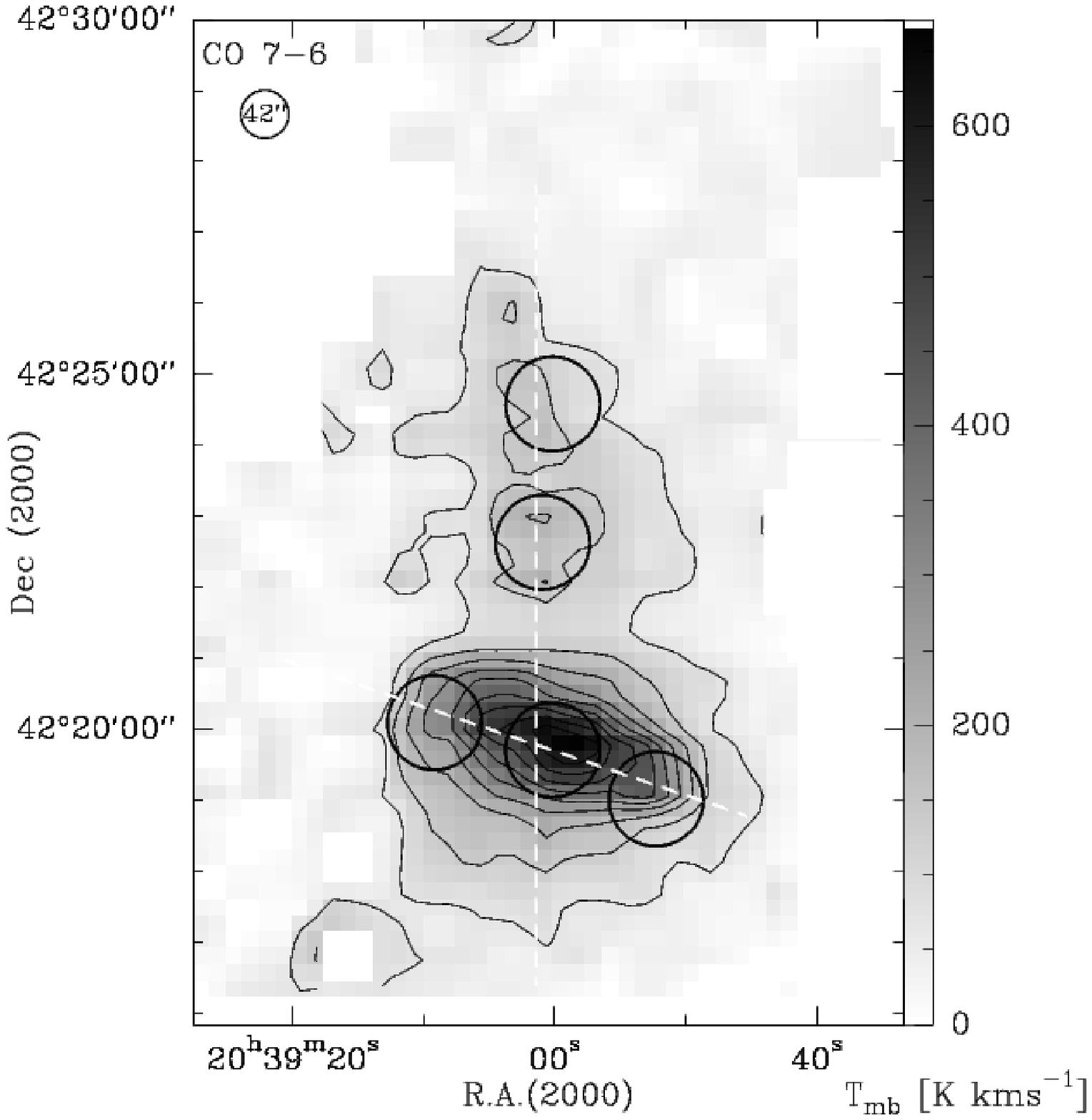}
\caption{Integrated intensity map of CO 4\TO3 ({\it left}) and CO 7\TO6 ({\it right}) of DR\,21 \& DR21\,(OH).
The integration interval is $-20$ to 30 $\kms$.
The contours for CO 4\TO3 run from 66($3\sigma$) to 462$\,\kkms$ in steps of $3\sigma$, and
from 68($2\sigma$) to 671$\,\kkms$ in steps of $2\sigma$ for CO 7\TO6.
The HPBW is indicated in the upper left corner.
The positions of ISO/LWS observations (cf. Fig. \ref{fig_spitzer}) are indicated by circles.
The dashed white lines correspond to the cuts shown in Fig. \ref{plot-co}.
\label{fig_map_co}
}
\end{figure*}
\begin{figure*}[ht]
\includegraphics[width=9.05cm,angle=-90]{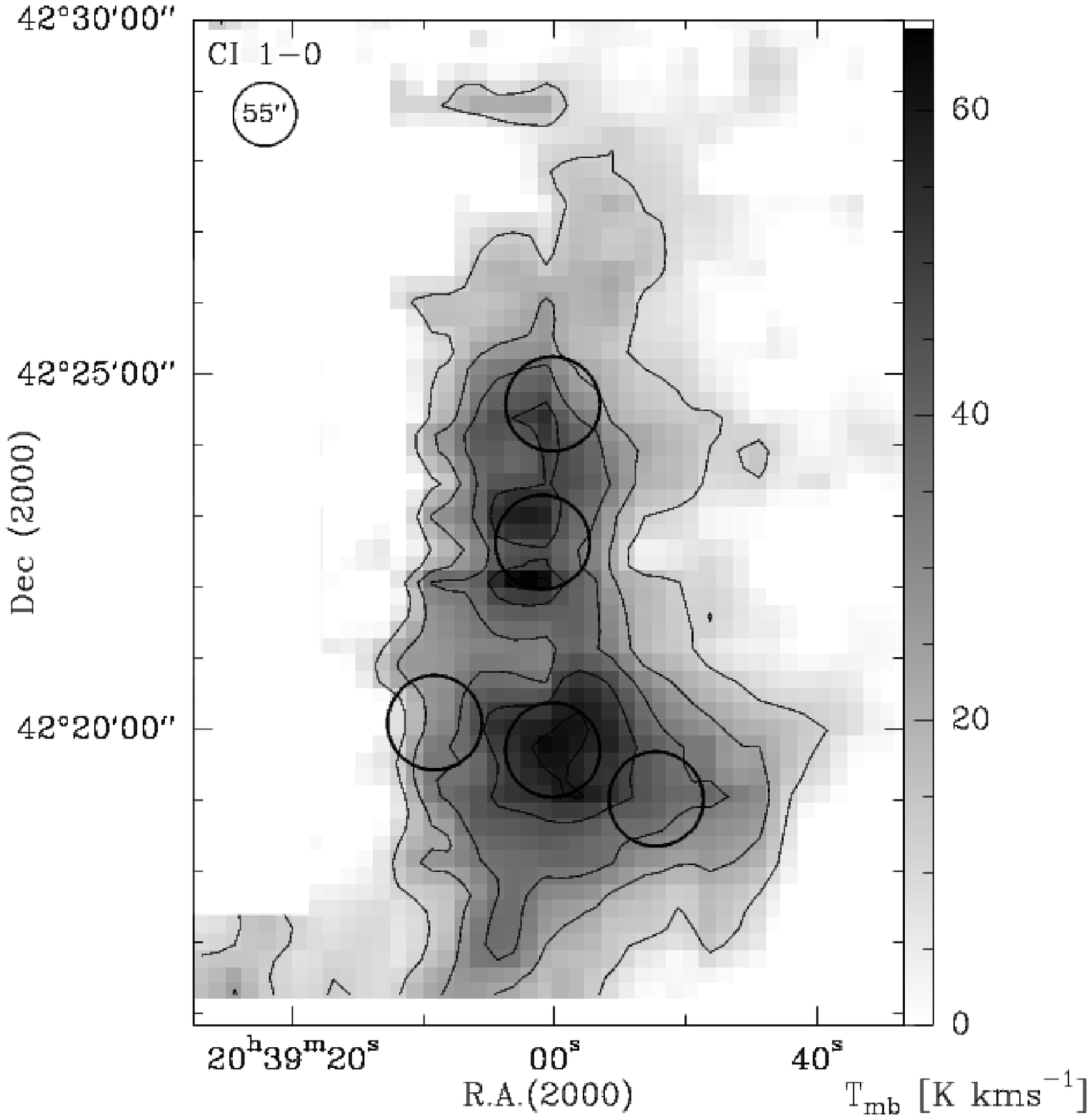}
\includegraphics[width=9.05cm,angle=-90]{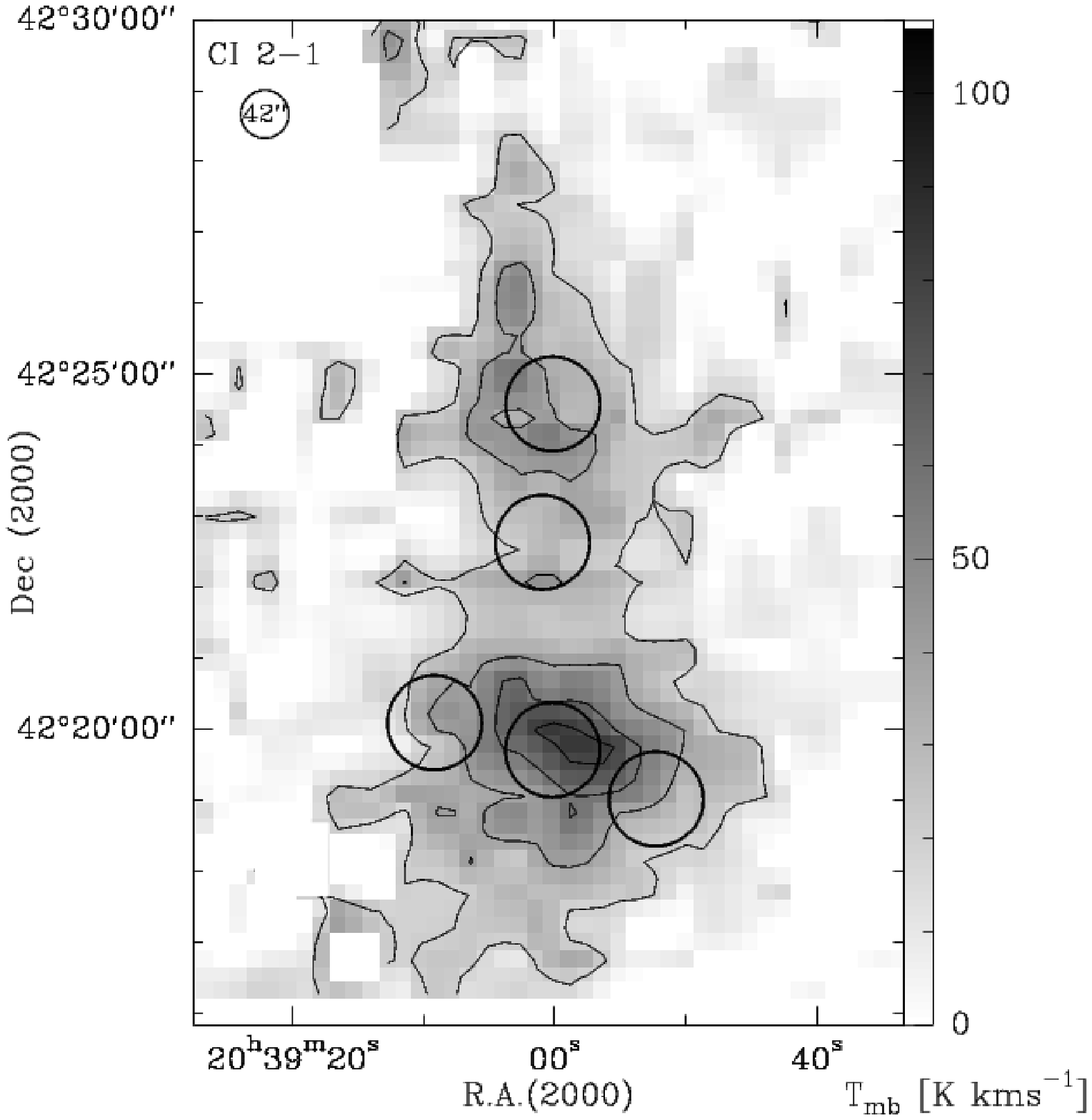}
\caption{Integrated intensity map of  \CI{} 1\TO0 ({\it left}) and of \CI{} 2\TO1 ({\it right}).
The integration interval is $-10$ to 20 $\kms$.
The contours for \CIw{} 1\TO0 run from 12($2\sigma$) to 60$\,\kkms$ in steps of $2\sigma$, and
from 20($1\sigma$) to 108$\,\kkms$ in steps of $\sigma$ for \CI{} 2\TO1.
\label{fig_map_ci}
}
\end{figure*}

%

\section{Observations} 

\OUT{
We used the KOSMA-3m submm-telescope to map the DR21/DR21\,(OH) region
in the CO 3\TO2, 4\TO3, 6\TO5, 7\TO6, $^{13}$CO 3\TO2, 6\TO5, and \CIw{}
1\TO0 and 2\TO1 transitions.
These data are combined with ISO/LWS FIR line observations of
the following transitions: \OI{} at 63, and 145\,$\mu$m, \CII{} at 158
$\mu$m, and the CO 14\TO13 to 17\TO16 high-$J$ CO transitions. The dataset is complemented
with $^{13}$CO and C$^{18}$O 1\TO0 FCRAO data toward single positions.
}

\subsection{KOSMA}

The region centered on DR21/DR21\,(OH)
was mapped in the two atomic
carbon fine structure transitions at 492\,GHz (609~$\mu$m,
$^3$P$_1\TO^3$P$_0$; hereafter $1\TO0$) and 809\,GHz (370~$\mu$m,
$^3$P$_2\TO^3$P$_1$; hereafter $2\TO1$) and the mid-$J$ rotational
transitions of CO ($J$=$3\TO2$, $J$=$4\TO3$, $J$=$6\TO5$, and $J$=$7\TO6$),
$^{13}$CO ($J$=$3\TO2$, and $J$=$6\TO5$)
using the dual channel 345 \& 230/660 GHz SIS receiver, and the
SubMillimeter Array Receiver for Two frequencies (SMART) at the KOSMA-3m telescope.
SMART is a dual-frequency, eight-pixel SIS-heterodyne receiver that
simultaneously observes in the 650 and 350\,\micron\ atmospheric
windows \citep{graf2002}.
{\NEW All observations were taken in the period between 2003 and 2006.}

An area of $\sim10'\times14'$ was mapped in the two [\ion{C}{i}] lines and the  CO 7\TO6 line on a
$27.5\arcsec\times 27.5\arcsec$ grid with a total integration time of 160\,s per position.
The observations before 2005 were performed in Dual-Beam-Switch mode (DBS), with a secondary mirror chop throw of $6'$ fixed in azimuth.
Due to the extent of the DR\,21 cloud complex, self-chopping is present in some of the DBS-mode observations.
In Appendix \ref{self-chopping}, we describe how we processed these data in order to reduce the impact of the self-chopping.
CO and $^{13}$CO $3\TO2$, and CO $4\TO3$ were observed in an On-The-Fly (OTF) mapping mode on a fully-sampled grid.
A zeroth order baseline was subtracted from the calibrated spectra.  
In addition to CO 3\TO2, we simultaneously obtained the CS 7\TO6 line in the image sideband of the 345\,GHz channel.
Pointing was frequently checked on
{\NEW
Jupiter,
the sun, and then confirmed on DR21 itself.
}
The resulting accuracy was better than $\pm20''$.
Larger $^{13}$CO 3\TO2 and 2\TO1 maps are presented in \citetalias{schneider2006}.
Due to their lower spatial resolution, we did not consider the $J$=2\TO1 data
here.

The $^{12}$CO and $^{13}$CO $6\TO5$ observations were
presented in \citet{koester1998}.
The typical DSB receiver temperatures in the high frequency channel was between 160--220\,K.
The $^{13}$CO $6\TO5$ observations cover a region of $\sim3'\times3'$
centered on DR\,21 while the $^{12}$CO observations extend further to the
north covering a region of $\sim3'\times6'$.

Atmospheric calibration was done by measuring the atmospheric emission
at the reference position to derive the opacity.
Sideband imbalances were corrected using standard atmospheric models \citep{cernicharo1985}
{\NEW assuming a receiver sideband gain of 0.5}.
Spectra taken simultaneously in the two frequency bands near 492 and 810\,GHz
were calibrated with an averaged water vapor value.
The \CI{} 2\TO1 and CO $7\TO6$ lines, observed simultaneously in two
sidebands, were also corrected for sideband imbalance.
Their pointing is identical and the relative calibration error
between these two transitions is very small.
{\NEW
We scaled the antenna temperature data
to the $T_\RM{mb}$ scale
using main beam efficiencies
$\eta_\RM{eff}$
as listed in Table~\ref{tab-obslist}.
The HPBWs ($40''$ to $80''$) and $\eta_\RM{mb}$
were determined using continuum scans on Jupiter.
}
From observations of standard calibration sources, we estimate that the absolute calibration
is accurate to within $\sim15\%$.
{\NEW
Most spectra were taken under
excellent weather conditions with typical values of the atmospheric opacity along the line-of-sight
better than 1.0 at 350 and 600\,\micron{}, and better than 0.5 above 870\,\micron{}.
}

\subsection{ ISO LWS Observations and Data Reduction}

ISO Long Wavelength Spectrometer (LWS, \citet{clegg1996}) $43-197$ $\mu$m grating 
scans (AOT L01) were obtained for 5 positions in the DR21 region from the ISO Data Archive (IDA)
(see Table \ref{tab-iso-positions} and Fig. \ref{fig_bb1}).
We removed duplicate scans, overlapping positions, or data with a very poor spectral resolution from the sample.

The grating scans contain the FIR continuum, the atomic fine structure lines of
\OI\ at 63 and 145\,\micron{}, ionised atomic lines of \CII{} at 158\,\micron{},
\OIII{} at 52 and 88\,\micron{}, \NII{} at 122\,\micron{}, 
and the high-$J$ CO molecular lines $J$=14\TO13 through $J$=17\TO16.
All spectra were processed with the ISO Spectral Analysis Package (ISAP, v. 2.1).
In ISAP, the data were deglitched by hand, defringed (detectors $4-9$), and corrected for flux clipping (the 
extended source correction). Some artefacts
remain after the processing
(see Fig. \ref{fig_bb1}), and were not corrected.
Line strengths were measured within ISAP. See Table \ref{tab-iso-pdr-lines} for further details.

\begin{figure}[tb]
\includegraphics[width=9.05cm,angle=-90]{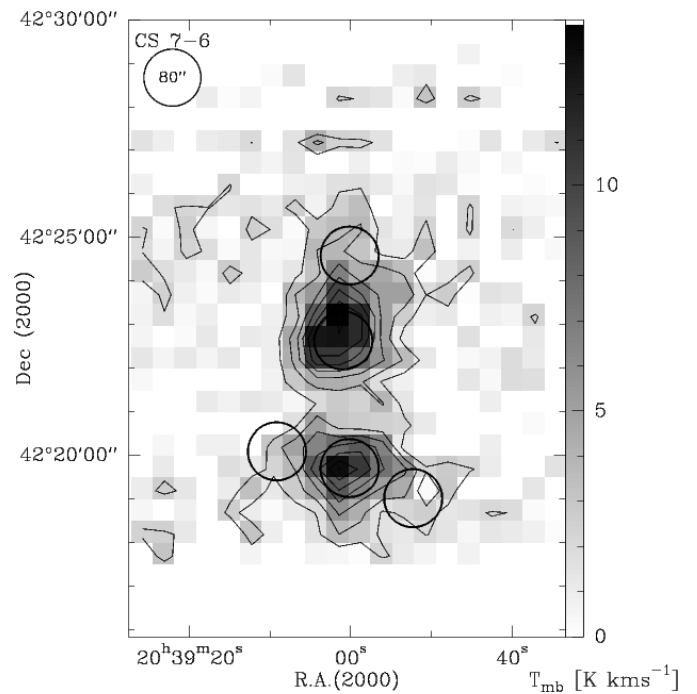}
\caption{Integrated intensity map of CS 7\TO6.
The integration interval is $-10$ to 4 $\kms$. Contours run from 1.8($3\sigma$) to 13.0$\,\kkms$ (in steps of $3\sigma$).
\label{fig_map_cs}
}
\end{figure}

%

\begin{figure*}
\includegraphics[width=0.6\linewidth,angle=-90]{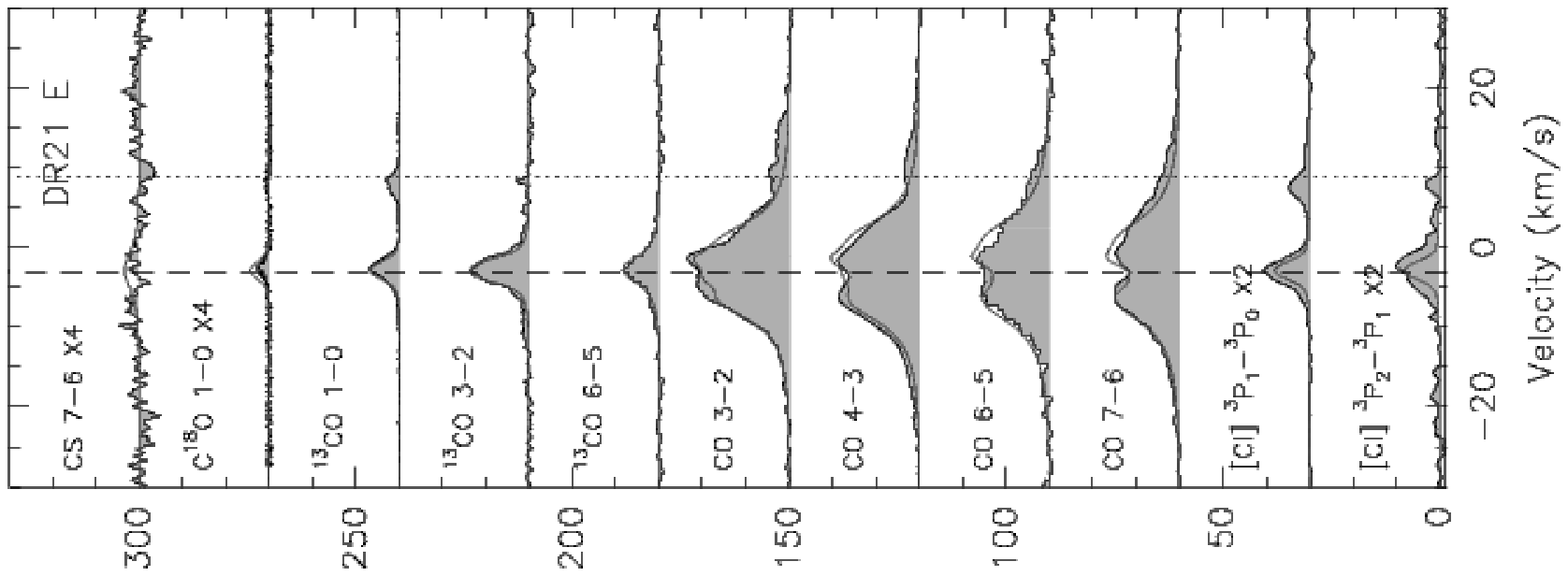}
\includegraphics[width=0.6\linewidth,angle=-90]{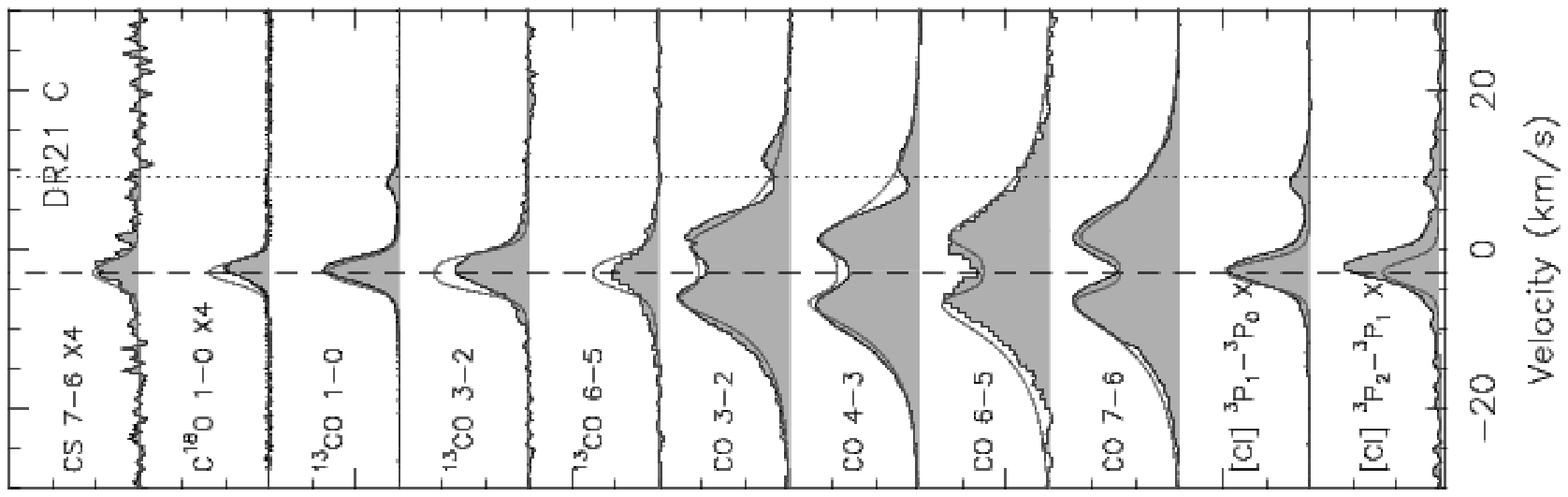}
\includegraphics[width=0.6\linewidth,angle=-90]{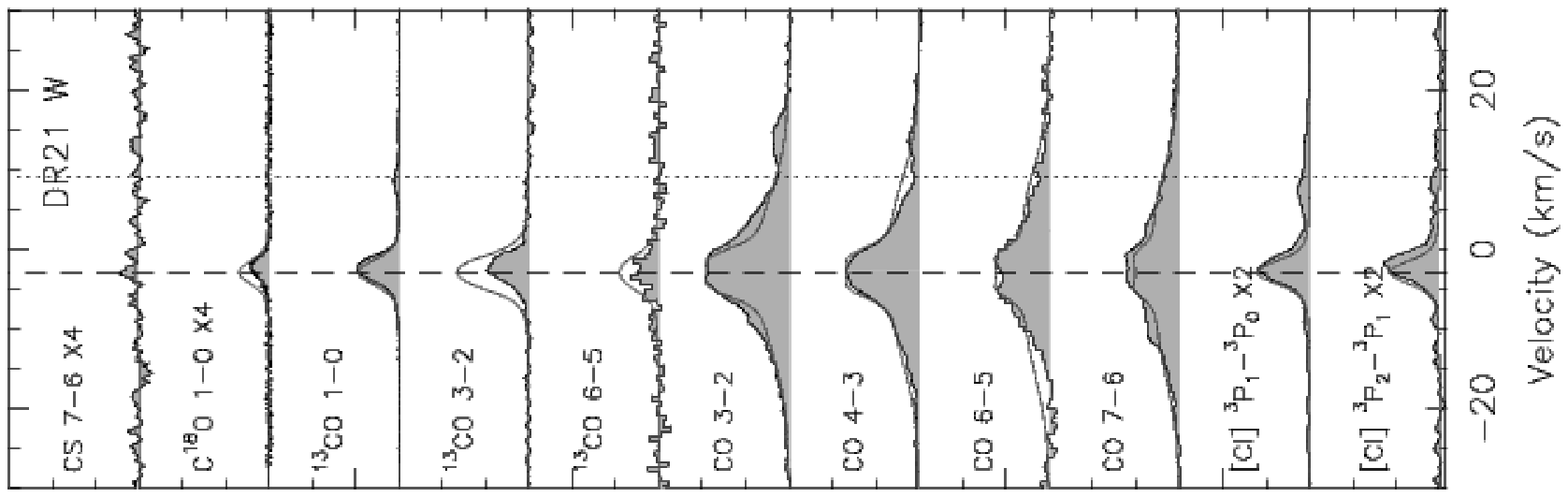}
\includegraphics[width=0.6\linewidth,angle=-90]{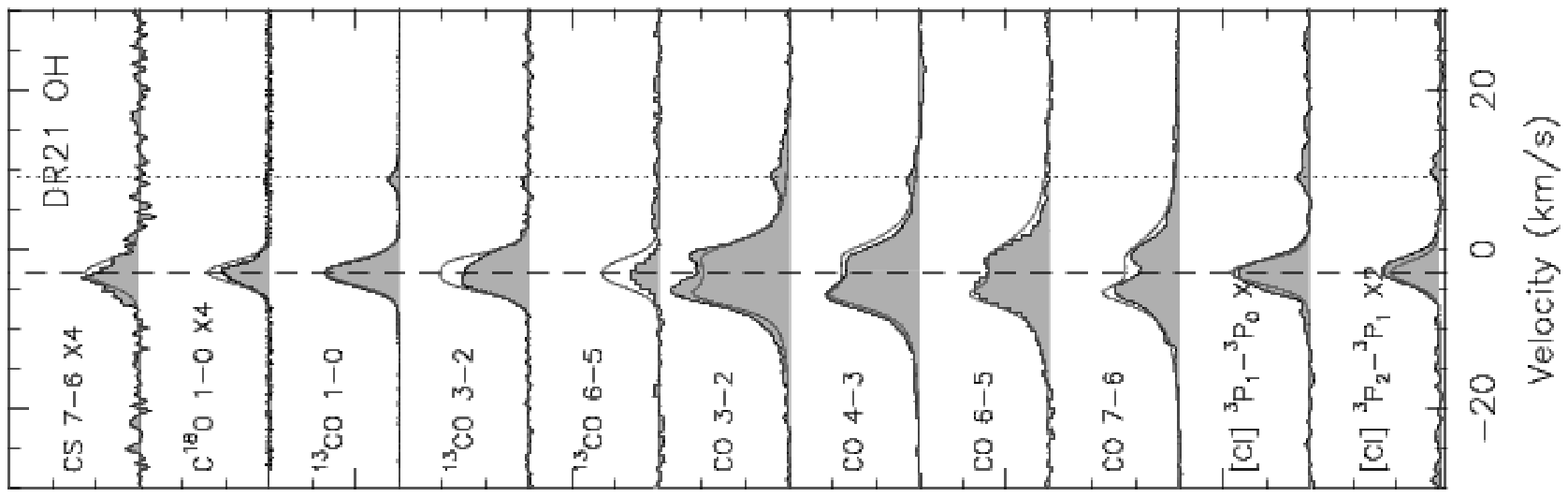}
\includegraphics[width=0.6\linewidth,angle=-90]{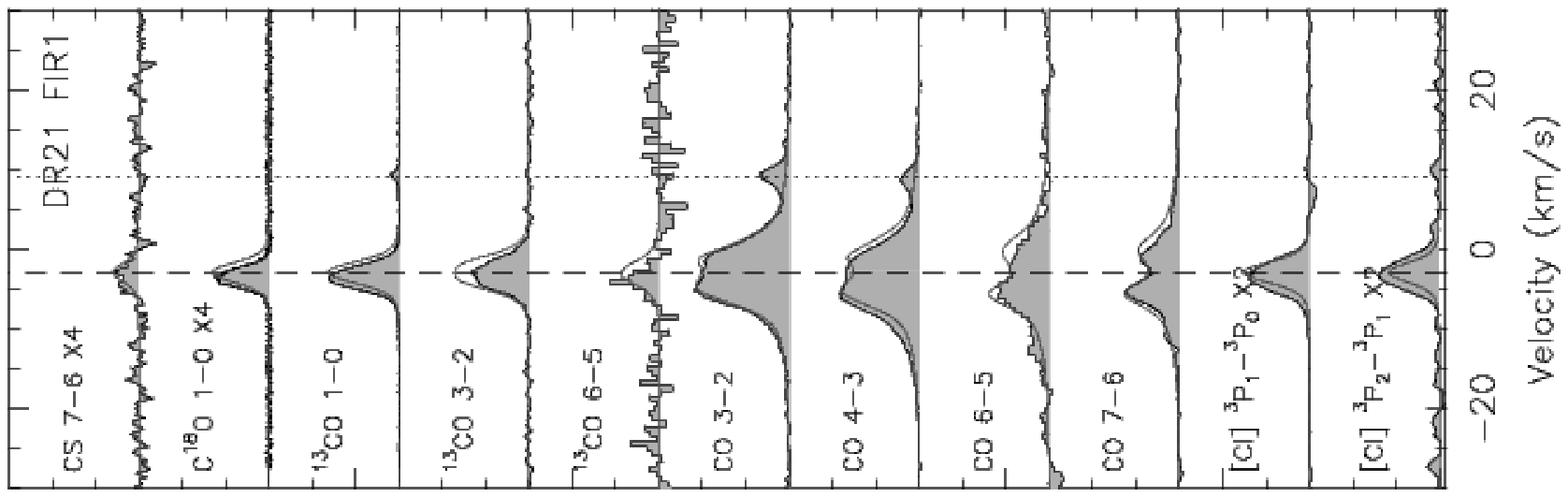}
\caption{Spectra of all observed species at the five ISO positions. {\it Grey
filled spectra outlined in black:} observed lines,
{\it thin grey lines:} line profiles modeled with the radiative transfer code {\sc SimLine} (see Sect. \ref{rad-model}).
All spectra are shown at a common spatial resolution of $80''$ and on a T$_\RM{mb}$ scale.
The baselines are shifted in steps of 30\,K.
\CI{} was multiplied by a factor of 2, CS 7\TO6 and C$^{18}$O 1\TO0 by a factor of 4.
$J=1\TO0$ transition data are FCRAO observations.
The dashed lines indicate the central velocity of $-3\,\kms$, and the dotted lines
the secondary velocity component at about $9\,\kms$, which was excluded from the fitting.}
\label{plot-spectra} 
\label{fig_bb2}
\end{figure*}

\subsection{FCRAO}

We use $^{13}$CO and C$^{18}$O 1$\TO$0 data from a large scale mapping
project of the whole Cygnus X region
using the FCRAO 14m
telescope. These data were obtained between 2003 December and
2006 January and will be presented in more detail in upcoming
papers (Simon et al., in prep.).
The data were observed using the 32 pixel array SEQUOIA in an On-The-Fly mapping mode
(at an angular resolution of $\sim$50$''$).
In this paper, we use smoothed data to have the same angular resolution of
$80''$ as the KOSMA data.

%

%

\section{Results}

Figure~\ref{fig_spitzer} gives an overview centered on DR21\,(OH) (designated as W75\,S by some authors).
The Spitzer 8\,\micron{} and 4.5\,\micron{} data by \citet{marston2004} show the small scale structure of this region at 
angular resolutions between $1.3''$--$3''$.
Several bright IR objects, in particular near DR21\,C{}\footnote{We will use the ISO/LWS position names troughout the paper (cf. Table \ref{tab-iso-positions}).},
delineate the dense ridge of gas and dust in both wavelength bands.
The map of 8\,\micron{} emission is dominated by many streamers of emission, which appear to connect to the ridge, and diffuse emission.
This is mainly emission from Polycyclic Aromatic Hydrocarbons (PAHs), which are the prime absorbers of far-ultraviolet (FUV) radiation from the surrounding and
embedded massive stars. Bright 8\,\micron{} emission therefore highlights the photon-dominated surface regions
of molecular clouds (PDRs).
The molecular outflow along DR21\,E{}, C{}, and W{} is well traced by $4.5$\,\micron{} emission, mainly due to
strong lines of shock excited or UV-pumped H$_2$ $\nu$$=$0\TO0 S(9).
No ro-vibrational lines of CO $\nu=$1\TO0 and only weak Br$\alpha$ have been found in this band \citep{Smith:2006dc}.

%

\begin{table*}[ht]
\caption[]{
Line integrated intensities in [$\kkms$] (upper table) and  in [$10^{-7}$ erg\,s$^{-1}$\,cm$^{-2}$\,sr$^{-1}$] (lower table) of the low- and mid-$J$  CO lines observed
(including the $^{13}$CO and C$^{18}$O isotopomers) with KOSMA and FCRAO, and CS 7\TO6.
In order to prevent the $+9\,\kms$ velocity component from adding to the integrated intensity,
the listed intensities were integrated only over the $-3\,\kms$ component.
As the $^{12}$CO lines merge with the second velocity component, these spectra were fitted using a 2-component gaussian
with the $+9\,\kms$ velocity component masked out.
All line intensities are reported for $80''$ angular resolution (higher angular resolution data have been smoothed).
\label{tab-pdr-lines}}
\begin{center}
\begin{tabular}{lrrrrrrrrrrrr}
\hline
\hline
\noalign{\smallskip}
Position&CO&CO&CO&CO&$^{13}$CO&$^{13}$CO&$^{13}$CO&C$^{18}$O&\CI&\CI&CS\\
Units in [$\kkms$]&3\TO2&4\TO3&6\TO5&7\TO6&1\TO0&3\TO2&6\TO5&$1\TO0$&1\TO0&2\TO1&7\TO6\\
\noalign{\smallskip}
\hline
\noalign{\smallskip}
DR\,21 E{} & 348.7& 283.4& 247.2& 265.7&22.38& 69.4& 43.1&1.66& 21.9& 28.2& 3.0\\
DR\,21 C{} & 440.0& 462.6& 434.4& 477.5&70.48&112.2& 80.9&8.26& 46.6& 58.4&13.8\\
DR\,21 W{} & 230.5& 221.9& 177.7& 223.1&41.02& 47.6& 33.9&3.67& 26.9& 28.5& 2.9\\
DR\,21 (OH)& 283.2& 197.4& 162.0& 153.2&67.92& 91.5& 24.9&9.03& 39.5& 30.2&13.6\\
DR\,21 FIR1& 223.8& 193.2& 130.2& 107.3&59.49& 61.9& 26.0&9.16& 33.4& 33.0& 3.9\\
\noalign{\smallskip}
\hline
\hline
\noalign{\smallskip}
Position&CO&CO&CO&CO&$^{13}$CO&$^{13}$CO&$^{13}$CO&C$^{18}$O&\CI&\CI&CS\\
in [$10^{-7}$erg\,s$^{-1}$\,cm$^{-2}$\,sr$^{-1}$]&3\TO2&4\TO3&6\TO5&7\TO6&1\TO0&$3\TO2$&$6\TO5$&1\TO0&$1\TO0$&$2\TO1$&7\TO6\\
\noalign{\smallskip}
\hline
\noalign{\smallskip}
DR\,21 E{}     &147.8& 284.6& 837.6&1429.1&0.307&25.7&127.6&0.023&26.8&153.4& 1.3\\
DR\,21 C{}$^a$ &186.5& 464.6&1471.9&2568.6&0.967&41.5&239.8&0.112&56.9&317.3& 5.7\\
DR\,21 W{}     & 97.7& 222.9& 602.1&1200.2&0.563&17.6&100.3&0.050&32.5&154.7& 1.2\\
DR\,21 (OH)    &120.0& 198.3& 548.9& 824.1&0.932&33.9&73.8 &0.122&48.3&164.7& 5.6\\
DR\,21 FIR1    & 94.8& 194.0& 441.2& 577.0&0.816&22.9&77.0 &0.124&40.8&179.5& 1.6\\
\hline
\end{tabular}
\end{center}
\begin{list}{}{}
\item $^a$ \citet{boreiko1991} give CO ($^{13}$CO) 9\TO8 intensities of $2.3\pm0.08\cdot 10^{-4}$ ($0.45\pm0.1\cdot 10^{-4}$)
erg\,s$^{-1}$\,cm$^{-2}$\,sr$^{-1}$ for a $80''$ ($83''$) beam and  a telescope coupling efficiency of 0.6.
\end{list}
\end{table*}

\subsection{Line-integrated maps}

\OUT{
Figures \ref{fig_spitzer} --
\ref{fig_map_cs} present maps of velocity integrated CO, \CI{} and CS emission.
{\MOD
These maps show a north-south oriented ridge of emission.
The intensity peak located at position DR21\,C{}, as well as
DR\,(OH) and DR21\,FIR1 lie on this ridge.
The latter one is the southernmost source of a chain of three far infrared spots (cf. Fig.\ref{fig_spitzer}),
}
}


\subsubsection{CO and $^{13}$CO maps}
\label{comaps}

\begin{figure}
\includegraphics[width=0.64\linewidth,angle=-90]{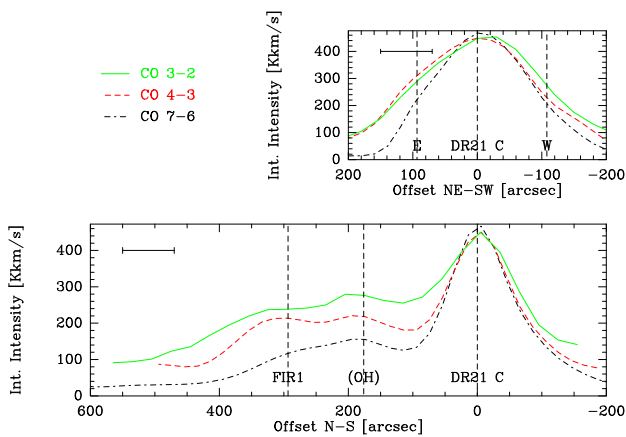}
\caption{Cut along the DR21 outflow axis {\it (top)}
and along the ridge {\it (bottom)} centered on DR21\,C{} in CO 3\TO2, 4\TO3, and 7\TO6 integrated emission.
The $80$\,\arcsec{} resolution is indicated as horizontal line. The dashed vertical lines indicate the positions
of the ISO/LWS observations.
}
\label{plot-co}
\end{figure}

Figure \ref{fig_spitzer} (contours on right panel) and Fig. \ref{fig_map_co}
show the KOSMA maps of integrated emission of the rotational transitions CO 3\TO2,
4\TO3, and 7\TO6 which preferentially trace the warm and dense molecular gas around DR21 and along the north-south orientated molecular ridge.
Apart from the ridge, the most obvious feature seen in the maps is the line emission
along the outflow axis oriented NE-SW.
The main axis of the strong 3\TO2, 4\TO3 and 7\TO6 emission (FWHM $\sim$230\,\arcsec{}, resp. $\sim 1.9$\,pc (D/1.7\,kpc)) is tilted by 20$^\circ$ with respect to
the right ascension axis and correlates nicely with extended H$_2$ emission lobes \citep{garden1986}.
The red-shifted CO line-wing emission is stronger toward the west side, whereas the blue-shifted emission is more pronounced to the east indicating
that this side is closer to the observer.
\OUT{
We will refer to the central position as DR21\,C{}, and to the east and west position of the outflow as DR21\,E{} and W{}.
}

Morphologically, the CO 3\TO2 map shows a very similar shape in comparison to the CO 4\TO3 data.
Detailed differences can be better seen if the maps are smoothed to the same resolution.
In Fig. \ref{plot-co}, we present two cuts along the outflow and along the main ridge from these data.
At DR21, all CO line integrated intensities are very similar.
Toward position DR21\,E, the emission of CO 3\TO2 and 4\TO3 stays at a high level while the 7\TO6 emission drops.
Toward the west, all three line intensities decline in a similar fashion.
\LIT{
\citet{jaffe1989} found CO 7\TO6 emission peaking on the \HII{}-region and following the H$_2$-emission.
They observed a peak flux of $1670\,\kkms$ (HPBW $32''$) at the 6 cm continuum peak \citep{dickel1983}
which is considerably higher than our peak of $756\,\kkms$.
We have no conclusive explanation for this disagreement.
}
Away from DR21, along the N-S cut, the CO 3\TO2 emission is dominant, whereas
CO 4\TO3 and CO 7\TO6 are sequentially weaker.
Strong CO 7\TO6 emission is limited to the DR21 core.
DR21\,(OH) is located $180$\,\arcsec{} north of DR21 and clearly separated from the DR21 region by a gap in the ridge.
DR21\,(OH) is bright in the lower $J$ transitions, but appears weaker in the $J$=7\TO6 transition.

$^{13}$CO 3\TO2 (contours on left panel of Fig. \ref{fig_spitzer}) was selected to map the column density in the region.
This transition traces the molecular ridge showing two maxima of emission at DR21\,C{} and
DR21\,(OH).
The total mass of the ridge is $3.5\times$10$^4$\,\msol (D/1.7\,kpc)$^2$ \citepalias{schneider2006}.
$^{13}$CO
weakly follows the outflow from DR21\,E{} to W{}.
Following the molecular ridge further to the north, an elongated emission band roughly
coincides in position with three far-infrared sources.
The southernmost source, DR21\,FIR1, was observed by ISO/LWS.
\citet{marston2004} find an extremly red object (ERO\,3) here, illuminating a shell-like halo in the 4.5\,\micron{} band.
The other two, FIR2 and FIR3,
coincide with MSX point sources.
As the FIR sources, this northern region is slightly tilted to the east, indicating flux contribution to mainly low-$J$ CO emission
from all three embedded far-IR sources.
$^{13}$CO reaches up to FIR3 which marks the edge of the dense ridge to the north.


\subsubsection{\CI{} maps}

\CIw{} 1\TO0 and 2\TO1 appear in good agreement with the shape of the molecular ridge as traced by $^{13}$CO.
Even though the emission around DR21 in \CI{} (see Fig.~\ref{fig_map_ci}) appears more centrally peaked than in the CO maps, the main axis between positions E{} and W{} is clearly visible.
The molecular ridge is seen to extend from the south east through DR21 and through DR21\,(OH)
$3'$ northwards.
DR21\,(OH) is visible as elongated, less compact region linking-up to the far-infrared sources further north.
The DR21\,(OH) region appears to harbour multiple sources as there is no central emission peak.
The \CI{} 1\TO0 map reveals a southern peak and extended weaker emission in the north.
This substructure may not be significant because the variation is only about $2\sigma$.
\OUT{, however, it agrees
with observations by \citet{mangum1992} and \citet{richardson1994}, who see four distinct cores (3 northern and 1 southern) in complementary tracers
such as NH$_3$ (at $4$\arcsec{} resolution), CS, and CH$_3$OH (about $20$\arcsec{}).
}
\citet{chandler1993} identified five compact sources in the DR21\,(OH) region (indicated in Fig. \ref{fig_spitzer}) using 1.3-mm thermal dust emission at $11''$ resolution.
\CIw{} 2\TO1 is, compared to 1\TO0, relatively weak toward DR21\,(OH) and suggests a lower temperature
than for DR21\,C{}
(see below).
North of DR21\,(OH), the carbon emission stretches over the three far-IR spots FIR1, FIR2, and FIR3.
Here, the upper atomic carbon line is in good agreement with the overall shape of optically thin low-$J$ CO isotopomers,
i.e., the C$^{18}$O 2\TO1 map by \citet{wilson1990}, and to some extent our $^{13}$CO 3\TO2 map.
A more distinct morphology analysis is not possible due to the remaining noise and some self-chopping residuals.

\subsubsection{CS 7\TO6 map}
\label{csmap}

The map in Fig. \ref{fig_map_cs} reveals two condensations at DR21 and around DR21\,(OH).
The southern peak is centered on the \HII{}-region and follows the direction of the outflow to the west.
The FWHM is $133$\,\arcsec{}$\times$$102$\,\arcsec{}.
The northern complex (FWHM  $108$\,\arcsec{}$\times$$156$\,\arcsec{})
is extended and may contain several unresolved cores (see above).
The emission extends up to the FIR1 position, but does not cover all three FIR spots.
The lines of CS 7\TO6 and C$^{18}$O 1\TO0
are the only lines in our sample that show brighter
emission toward DR21\,(OH) than toward the southern molecular cloud/\HII{}-region DR21.
For CS, this has also been observed by \citet{shirley2003}.

\subsection{Line profiles}

\label{specta_pos}

In the following, we concentrate on the 5 ISO/LWS positions (as indicated in Fig.~\ref{fig_spitzer}).
In Fig. \ref{plot-spectra}, we present spectra of $^{12}$CO, $^{13}$CO, and C$^{18}$O towards these positions together with spectra of \CIw{}, and CS 7\TO6 between $-30$ to $30\,\kms$.
We present all observations at a common resolution of $80\arcsec$.
In Table~\ref{tab-pdr-lines}, we give the corresponding line fluxes.

The $^{12}$CO emission from the DR21/\HII-region shows broad wing emission due to the outflowing molecular gas
(with line widths of more than $15\,\kms$ and peak temperatures well above 20\,K).
A self-absorption dip at the velocity of $-3\,\kms$ is most pronounced in mid-$J$ CO lines at DR21\,C{} where the peak intensity drops by a factor of two at the line center.
Toward DR21\,(OH), CO shows blue-skewed, double-peaked lines with self-absorption. 
In both cases, the velocity of the line reversal nicely matches the corresponding emission in $^{13}$CO and \CIw{}, and is likely
caused by the same cold material in the dust lane.
\LIT{
\citet{jaffe1989} identified a very deep ($<5$\,K) and narrow central reversal in CO 7\TO6 emission toward the DR21 peak (FWHM $\sim$$25\,\kms$),
implying a kinetic temperature gradient of $>100$\,K down to less than $20$\,K.
}

With widths of $4$ to $6\,\kms$, the line shapes of $^{13}$CO are similar to the \CI{} profiles.
The $^{13}$CO $6\TO5$ emission ($E_u/k_B=79.3$\,K) at all five positions indicates the presence of warm and dense gas.
\LIT{
as already pointed out by \citet{boreiko1991} who observed the CO and $^{13}$CO 9\TO8 lines at DR21\,C{} (see Table \ref{tab-pdr-lines}).
}
At DR21\,C{}, $^{13}$CO $6\TO5$ shows broad symmetric wing emission that is underlying a narrow component.
The velocity extent of the wings is similar to that in mid-$J$ CO lines.
A better signal-to-noise ratio would allow a more accurate estimate.
Only weak line wings are visible in the lower $^{13}$CO 3\TO2 line since it
traces primarily the quiescent gas.

DR21 is associated with molecular gas centered at a velocity of $-3\,\kms$.
In addition to this, weaker emission between $\sim 7$ and $14\,\kms$ is associated with the W75\,N complex
approximately $11$\,pc\,(D/1.7\,kpc) to the north-west.
Some authors \citep[e.g.,][]{dickel1978} conclude that both clouds are actually interacting. \citetalias{schneider2006} shows channel maps revealing a link between both clouds.
Diffuse emission in the $^{13}$CO $J=1\TO0$ and both \CI{} transitions, and
absorption in the red wing of the low-$J$ CO lines (up to $J=4\TO3$) suggest that this material is located in front of the DR21 complex.
We do not further discuss this contribution here and refer to \citetalias{schneider2006}.

%

\begin{table*}[ht]
\caption[]{
The ISO/LWS high-$J$ CO and atomic FIR lines at 80$''$ resolution.
The transformation between flux densities and surface brightnesses was done using the 
beam sizes  given in Table 5.10 of the ISO-LWS handbook v. 2.1.
The beam solid angle varies between $8\times10^{-8}$\,sr and $1.4\times 10^{-7}$\,sr.
Because the lines are unresolved, the line widths were held to the instrumental line width: 0.29 or 0.60 $\mu$m
for detectors $0-4$ and $5-9$, respectively.
Correction factors between 0.37 to 0.85 for extended sources have been applied.
We assume a 30\% absolute photometry uncertainty for LWS.
For DR21\,(OH), numbers in parentheses are more uncertain (S/N between $1-3$) due to a poorer frequency resolution.
At DR21\,FIR1, no detection of CO was possible, so we give $3\sigma$ upper limits.
\NII{} was clearly found only toward the outflow positions E{} and W{}. At the other positions we give $3\sigma$ upper limits.
}\label{tab-iso-pdr-lines}
\begin{tabular}{lllllllrrrrrrrrr}
\noalign{\smallskip} \hline \hline \noalign{\smallskip}
Position&CO&CO&CO&CO&\CII&\OI&\OI&\OIII&\OIII&\NII\\
in [$10^{-4}$erg\,s$^{-1}$\,cm$^{-2}$\,sr$^{-1}$]&14\TO13&15\TO14&16\TO15&17\TO16&158\,\micron&63\,\micron&145\,\micron&52\,\micron&88\,\micron&122\,\micron{}\\
\noalign{\smallskip} \hline \noalign{\smallskip}
DR\,21 E{} & 0.25 & 0.72 & 0.51   & 0.33 & 8.2 & 48.2 & 2.9  &3.53&3.08&0.435\\
DR\,21 C{} & 1.75 & 2.46 & 0.99   & 0.63 & 8.1 & 49.4 & 6.15 &11.5&4.64&$<0.75$\\
DR\,21 W{} & 0.44 & 0.69 & 0.42   & 0.29 & 4.7 & 10.1 & 1.136&1.78&2.52&0.306\\
DR\,21 (OH)& (0.45)& (0.99)&(0.40)&(0.50)& 2.3 & 2.94 & 0.845&(1.27)&1.56&$<2.7$\\
DR\,21 FIR1&$<0.28$&$<0.3 $&$<0.28$&$<0.25$&3.6&1.78& 0.45 &0.54&1.66&$<0.6$\\
\hline
\end{tabular}
\end{table*}

%

\begin{table*}[ht]
\caption[]{Results of dust continuum single-component fit.
$\beta$ was fixed at $1.5$, and $\Omega$ is $1.5\cdot 10^{-7}$\,sr.
The extinction and the opacity are linked through $\RM{A}_\RM{v}=2.5 \tau_\RM{v} \log_{10} e\approx 1.085 \tau_\RM{v}$.
The total infrared flux is calculated from $I_\RM{IR}=\int F_\lambda d\lambda$.
The luminosity $L$ and mass $M$ are given for 1.7\,kpc distance.
}\label{iso-dust}
\begin{tabular}{lcccccccccc}
\noalign{\smallskip} \hline \hline \noalign{\smallskip}
  Position &$\tau_\mathrm{v}$&A$_\mathrm{v}$(ISO)&N(H$_2$)&Mass&T$_\mathrm{dust}$&$I_\RM{IR}$&$L$\\
 & &&[$10^{22}$\,\cmsq]&[\msol{}(D/1.7\,kpc)$^2$]&[K]&[erg\,s$^{-1}$\,cm$^{-2}$\,sr$^{-1}$]&[$10^3L_\odot$(D/1.7\,kpc)$^2$]\\
\noalign{\smallskip} \hline \noalign{\smallskip}
DR\,21 E{} & 20&$21.2\pm1.9$& 2.1 &148 &$41.6\pm0.6$& $0.55 \pm0.05$ & 7 \\
DR\,21 C{} &112&$121 \pm11$&12.1 &850 &$42.5\pm0.7$& $3.4  \pm0.3 $  & 45\\
DR\,21 W{} & 34&$36.9\pm3.5$&3.69 &259 &$30.7\pm0.4$& $0.18\pm0.02$  & 2 \\
DR\,21 (OH)&238&$259\pm43$&25.8 &1814&$31.2\pm0.8$& $1.3 \pm0.2 $    & 17\\
DR\,21 FIR1& 45&$48.4\pm4.3$&4.84 & 340&$33.3\pm0.5$& $0.37\pm0.03$  & 5 \\
\noalign{\smallskip} \hline
\end{tabular}
\end{table*}

%

\subsection{ISO archive data}
\label{dust-section}

\begin{figure}[htb]
\includegraphics[height=0.94\linewidth,angle=-90]{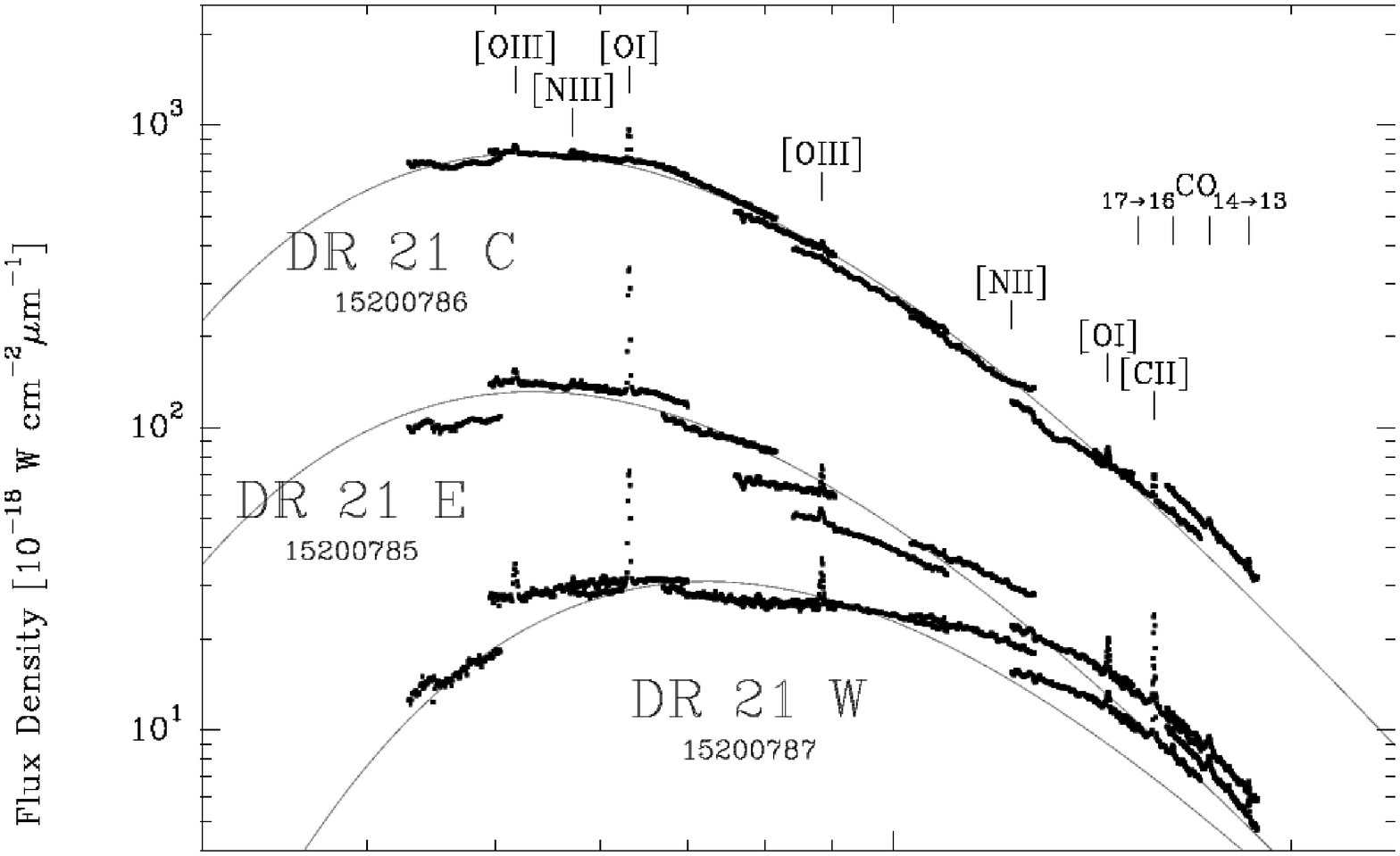}\break
\includegraphics[height=0.94\linewidth,angle=-90]{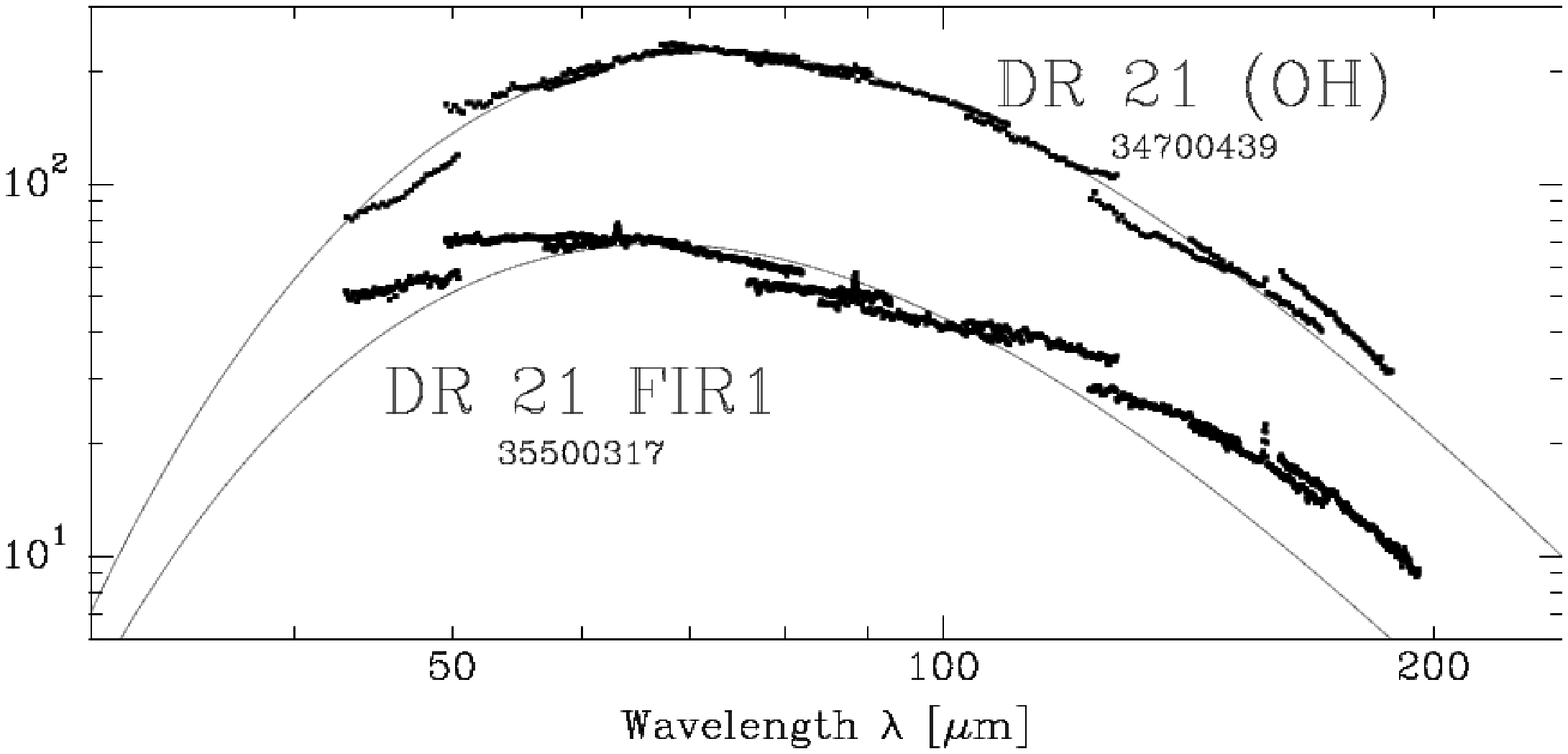}
\caption{
ISO/LWS Spectral Energy Distribution (SED) of the DR\,21 region {\it (upper panel)}, and
the DR21\,(OH) and DR21\,FIR1 region {\it (lower panel)}. Curves indicate the best greybody
fit corresponding to the values given in Table~\ref{iso-dust}.}
\label{fig_bb1}
\end{figure}

\label{chap-spectra}

Figure~\ref{fig_bb1} shows the ISO/LWS continuum fluxes together with spectrally unresolved lines
of \CII{}, \OI{}, \OIII{}, \NII{}, and rotational high-$J$ CO transitons.
The following Sections discuss the dust and line emission at the individual positions.

\OUT{
\begin{table}[ht]
\caption[]{Results of dust continuum single-component fit.
$\beta$ was fixed at $1.5$, and $\Omega$ is $1.5\cdot 10^{-7}$\,sr.
The extinction and the opacity are linked through $\RM{A}_\RM{v}=2.5 \tau_\RM{v} \log_{10} e\approx 1.085 \tau_\RM{v}$.
The total infrared flux (TIR) is calculated from $I_\RM{IR}=\int F_\lambda d\lambda$.
The luminosity $L$ and mass $M$ are given for 1.7\,kpc distance.
}\label{iso-dust2}
\begin{tabular}{lcccccccccc}
\noalign{\smallskip} \hline \hline \noalign{\smallskip}
  Position &$\tau_\mathrm{v}$&A$_\mathrm{v}$(ISO)&$I_\RM{IR}$\\
 & &&[erg s$^{-1}$ cm$^{-2}$ sr$^{-1}$]\\
\noalign{\smallskip} \hline \noalign{\smallskip}
DR\,21 E{} & 20&$21.2\pm1.9$&$0.55 \pm0.05$\\
DR\,21 C{} &112&$121 \pm11$ & $3.4 \pm0.3$\\
DR\,21 W{} & 34&$36.9\pm3.5$&$0.18\pm0.02$\\
DR\,21 (OH)&238&$259\pm43$&  $1.3 \pm0.2 $\\
DR\,21 FIR1& 45&$48.4\pm4.3$&$0.37\pm0.03$\\
\noalign{\smallskip} \hline
\end{tabular}
\end{table}
}

\subsubsection{Grey body fit to the continuum}
\label{greybodyfit}

In all five cases, the far-infrared dust-continuum flux -- with strong emission lines masked out -- was $\chi^2$-fitted with a single isothermal grey-body model
$F(\RM{T}_\RM{dust},\tau_\RM{dust})_\lambda=\Omega B_\lambda(T_\RM{dust})(1-e^{-\tau_\RM{dust}})$ with
fixed solid angle ($\Omega=1.5\cdot 10^{-7}$\,sr, assuming that the emission fills the $80''$ aperture)
and a dust opacity $\tau_\RM{dust}\sim\tau_\RM{v}\lambda^{-\beta}$ with $\beta=1.5$.
Spectral indices are typically found between $1.0$ and $2.0$ (see \citet{goldsmith1997} for a discussion of observations and grain models).
Although the spectral index is known to vary with the cloud type and optical depth, we assume a fixed $\beta$ for all 5 positions (but see discussion in Sect. \ref{massfilling}).
A fixed $\beta$ of 1.5 is compatible with grain models of \citet[][]{preibisch1993}, \citet{pollack1994}, and observations,
e.g., by \citet[][]{walker1990}.
The total infrared dust continuum flux and bolometric luminosity per position is derived by integrating the fit-model $F_\lambda$.
The integration interval ($1-1000$\,\micron{}) covers the expected intensity maxima between 50--80\,\micron{} and corresponds to the range used,
e.g., by \citet{calzetti2000}.
Errors of the fit parameters were estimated as described in the following:
we assume
{\it (i)} the optical depths $\tau_\RM{dust}$ and total fluxes are accurate to about $\sim10$\,\%, mainly due to the remaining calibration artefacts within the LW1$-$5 bands.
{\it (ii)} the temperature fit error is less than  2\,\% but the absolute error also depends on the quality of SW1 detector data (leftmost data points in Figure~\ref{fig_bb1}).

We find extinctions between $20^m$$\le$$\RM{A}_v$$\le$$260^m$ and temperatures between $31$$-$$43$\,K (cf. Table~\ref{iso-dust}).
In contrast to the extinction, the variation in dust temperature is small.
Considering that we use a single-temperature model, these temperatures are probably biased by the warmer dust.
Evidence of cold dust peaking above 100\,\micron{} is visible at DR21\,W{}, E{}, (OH) and FIR1 in Fig. \ref{fig_bb1} but
disentangling this component would require longer wavelength data
and the contribution to the integrated grey-body flux is relatively small.
The cold dust is studied in detail by, e.g., \citet{chandler1993} and \citet{motte2005}.
We derive the molecular hydrogen column density from the conversion N(H$_2$)/A$_v\approx 0.94\cdot 10^{21}$\,cm$^{-2}$\,mag$^{-1}$ \citep[][ cf. Table \ref{iso-dust}]{bohlin1978}
and if placed at the distance of 1.7\,kpc, the molecular gas masses vary between $150-1800$\,\msol{}.

The peak of the beam-averaged dust temperature is $\sim$43\,K at DR21\,C{}.
With A$_v>100^m$, the strongest continuum sources are DR21\,C{} and DR21\,(OH).
Due to the high temperature towards DR21\,C{}, the flux density is higher at short wavelengths,
but drops below the flux density of DR21\,(OH) above $\sim$$200$\,\micron{}.
For DR21\,(OH), we estimate a higher mass than for the DR21\,E{}--W{} complex (1814, resp. 1257\,\msol{}).
\citet{chandler1993} assumed temperatures between 25\,K and 40\,K and
derived a slightly lower total mass of 1650\,\msol{}$(D/1.7\,\RM{kpc})^2$ for the sources at DR21\,(OH).
\OUT{
Their total luminosity is lower by a factor of 3 compared to the ISO results.
At the position FIR1, which covers the more massive FIR2 in the ISO beam, we derive in contrast to \citeauthor{chandler1993}
a mass higher by a factor of two.
Because our result of $\sim 31$\,K is right within their assumed temperature range,
the warm dust that is likely associated with FIR1 must contribute to our temperature estimate
on the larger scale of the ISO beam.
}

%

\subsubsection{ISO/LWS far-IR lines: Carbon Monoxide and Ionised Carbon}
\label{isocolines}

The LWS band covers many important far-IR cooling lines and
tracers of the photon dominated regime of molecular clouds.
Table~\ref{tab-iso-pdr-lines} gives the integrated line intensities of \CII{} (158\,\micron{}), high-$J$ CO, and \OI{} (63 and 145\,\micron{})
derived from the ISO observations quoted above.
In this paragraph, we focus on the carbon bearing species
needed for Sect. \ref{dr21phys} and postpone the discussion of the other lines to the end of Sect. \ref{dr21phys}.

At DR21, highly excited CO emission is visible at all three positions, but most prominently around the \HII-region at DR21\,C{}
The center position with more than $2.4\cdot 10^{-4}$\,erg\,s$^{-1}$\,cm$^{-2}$\,sr$^{-1}$ in mid- and high-$J$ CO
is also the brightest continuum source in the mapped region.
The emission stems from large columns of warm quiescent gas that is heated from an embedded object.
Toward the East and West positions, where vibrationally excited H$_2$ outlines the molecular outflow \citep{garden1986},
the FIR continuum is somewhat weaker and more importantly, the continuum flux does not correlate with the unusually
strong FIR line emission.
This gas is visible in the line wings of mid-$J$ CO lines (Fig. \ref{plot-spectra}) and in bright H$_2$ line emission
at 4.5\,\micron{} in the IRAC map (Fig.~\ref{fig_spitzer}).
\citet{lane1990} conclude that neither a single C- nor a single J-shock can account for the observed far-IR line emission alone and
additional heating through FUV radiation is required.
\OUT{
(a detailed discussion of the UV flux is presented in Appendix \ref{uvfield})
}
Although DR21\,(OH) has the second strongest continuum, the sources associated with this position are surprisingly inactive in far-IR line emission.
We find four high-$J$ CO line detections, but due to a lower scanning resolution, the values given
in Table \ref{tab-iso-pdr-lines} have higher uncertainty.
The far-IR source FIR1 is the faintest with respect to the line fluxes, although the continuum is stronger than towards DR21\,W{}.
High-$J$ CO was not detected at FIR1, so we give upper limits in the table.

\CII{} is detected at all five positions.
The emission peaks at DR21\,E{} and toward the West, the flux decreases by almost 50\,\%.
The low \CII{} flux of $2.3\cdot 10^{-4}$\,erg\,s$^{-1}$\,cm$^{-2}$\,sr$^{-1}$
at DR21\,(OH) may indicate a lack of ionised carbon or
that the temperature is too low for excitation in the upper energy level
($\RM{E}_u/k_B=91$\,K above ground).
At DR21\,FIR1, \CII{} is again slightly enhanced compared to DR21\,(OH).


\section{The physical structure of DR\,21}
\label{dr21phys}

In this section we analyse the line properties of the observed carbon species from ISO/LWS data
and combine these results with the KOSMA observations.
We start with an LTE approach and then model the line emission with full radiative transfer.

%

\subsection{Comparison of CO, C, and C$^+$}
\label{civsco}

We estimate the composition of the molecular gas with respect to the carbon abundance in the main gas-phase species CO, C, and C$^+$.
This is done by assuming LTE and optically thin line emission.
We therefore use the less abundant isotopomeres $^{13}$CO and C$^{18}$O.
We list the results for excitation temperature, column density and abundance in Table \ref{ci-table}.

\subsubsection{Atomic Carbon}
\label{atomiccarbon}
The \CIw{} excitation temperature needed to produce the observed line ratios can be estimated since both \CIw{} lines are likely to be optically thin
(cf. line profiles in Fig. \ref{plot-spectra}).
The line temperature ratio $R_\RM{\CIw{}}$ falls between $0.8$ at DR21\,(OH) and $1.3$ around and east of DR21.
The temperature is determined from this ratio using $T_\RM{ex}=38.8\,\RM{K}/\ln(2.11/R_\RM{\CIw{}})$ (Eq. (\ref{ciratio})).
The range (38\,K at (OH) to 79\,K at E{}) indicates the origin of the atomic carbon lines in a warm environment.
\citet{zmuidzinas1988} derived a temperature of 32\,K from the two \CIw{} lines at DR21\,(OH) which is close to our value at this position.
Our results for T$_\RM{ex}$ are lower compared to other Galactic high mass star forming regions
\citep[e.g., W3\,Main and S\,106 with typically T$_\RM{ex} \ge 110$\,K, ][]{kramer2004,schneider2003},
but substantially higher compared to the average temperature in the inner Galactic disk of $\sim 20$\,K \citep{fixsen1999}.
\OUT{A minumum kinetic temperature in M82, $50$\,K \citep{stutzki1997}, was estimated using the same method.}
{\NEW The beam averaged column density, $3.2-6.8\cdot 10^{17}$\,\cmsq{} (Eq. (\ref{columnci})),}
shows variations by a factor of 2 or less, peaking at the position DR21\,C{} and around (OH), i.e., towards the densest condensations along the ridge.
This is remarkable since the H$_2$ column density derived from the dust
was found to vary by a factor of 12.
The column density for DR21\,(OH) of $6.3\cdot 10^{17}$\,\cmsq{} as determined by \citeauthor{zmuidzinas1988} is only slightly higher than our result.

\begin{table*}[ht]
\caption[]{\CIw{} and $^{13}$CO line ratio $R$ (in $\kkms$; values taken from Table~\ref{tab-pdr-lines}),
excitation temperature (with lower and upper $1\sigma$ limits), and \CIw{}, $^{13}$CO (with relative upper and lower errors),
and C$^{18}$O column densities (bracketed by $T_\RM{ex}$ of $^{13}$CO).
The \CIIw{} column density was estimated using a method described in Sect. \ref{civsco}.
All abundances X are relative to the H$_2$ column densities (first line; taken from Col. 4 of Table \ref{iso-dust}).
The last row lists the relative abundance of carbon in the three major gas phase species \CIIw{}, \CIw{}, and CO.
The CO column density is derived from the lower C$^{18}$O column density assuming a relative [CO]:[C$^{18}$O] abundance of 490.
}\label{ci-table}
\begin{center}
\begin{tabular}{lrcccccc}
\noalign{\smallskip} \hline \hline \noalign{\smallskip}
Line/Lineratio&Source Property &DR21\,E{} & DR21\,C{} & DR21\,W{} & DR21\,(OH) & DR21\,FIR1\\
\noalign{\smallskip} \hline \noalign{\smallskip}
\noalign{\smallskip}
H$_\RM{2, dust}$&N(H$_2$) [$10^{22}$\,\cmsq]&2.1&12.1&3.69&25.8&4.84\\
\noalign{\smallskip}
\hline

\noalign{\smallskip}
\CIw{}&$R_\RM{\CIw{}}$ (\CIw{} 2\TO1/1\TO0) &$1.29\pm 0.36$ & $1.25\pm 0.35$& $1.06\pm 0.30$& $0.77 \pm 0.22$ &$ 0.99\pm 0.28$\\
&$T_\mathrm{ex,\CI}$ [K]&79$_{47}^{157}$ &74$_{50}^{140}$ & 56$_{41}^{88}$ & 38$_{31}^{51}$ & 51$_{39}^{76}$\\
&N(\CIw) [$10^{17}$\,\cmsq]& $3.2_{-0.2}^{+0.3}$ & $6.8_{-0.4}^{+0.6}$ & $3.8_{-0.2}^{+0.3}$ & $5.2_{-0.1}^{+0.2}$ & $4.6_{-0.2}^{+0.3}$\\
&X(\CIw) [$10^{-5}$]&1.5&0.56&1.03&0.20&0.10\\

\noalign{\smallskip}
\hline

\noalign{\smallskip}
$^{13}$CO 6\TO5/3\TO2&$R_{^{13}\RM{CO,high}}$ ($^{13}$CO 6\TO5/3\TO2)        &$0.62\pm0.26$&$0.72\pm0.31$&$0.71\pm0.30$&$0.27\pm0.12$&$0.42\pm0.18$\\
&$T_{\RM{ex},^{13}\RM{CO,high}}$ [K]&$43^{53}_{36}$&$46^{58}_{38}$&$46^{58}_{38}$&$30^{34}_{26}$&$35^{41}_{30}$\\
&N($^{13}$CO 6\TO5) [$10^{15}$\,\cmsq]&$33.0_{+11.1}^{-9.0}$&$53.8_{+20.4}^{-12.8}$&$22.6_{+8.5}^{-5.4}$&$39.6_{+21.3}^{-10.7}$&$27.9_{+13.4}^{-7.2}$\\
\noalign{\smallskip}
$^{13}$CO 3\TO2/1\TO0&$R_{^{13}\RM{CO,low}}$ ($^{13}$CO 3\TO2/1\TO0)        &$3.10\pm1.30$&$1.59\pm0.67$&$1.16\pm0.49$&$1.35.\pm0.57$&$1.04\pm0.44$\\
&$T_{\RM{ex},^{13}\RM{CO,low}}$ [K]&$25^{37}_{16}$&$15^{19}_{12}$&$13^{16}_{10}$&$14^{17}_{11}$&$12^{15}_{10}$\\
&N($^{13}$CO 1\TO0) [$10^{15}$\,\cmsq]&$33.4_{-8.9}^{+12.3}$&$75.6_{-8.6}^{+12.2}$&$40.0_{-4.6}^{+5.1}$&$69.1_{-8.0}^{+8.6}$&$56.5_{-4.3}^{+7.2}$\\
\noalign{\smallskip}
C$^{18}$O 1\TO0&N(C$^{18}$O) [$10^{15}$\,\cmsq]&$2.4-3.8_{}^{}$&$8.7-20.0_{}^{}$&$3.5-8.9_{}^{}$&$9.0-15.1_{}^{}$&$8.5-17.7_{}^{}$\\
&X(C$^{18}$O) [$10^{-7}$]&$1.2-1.8$&$0.72-1.7$&$0.95-2.4$&$0.35-0.59$&$1.7-3.7$\\


\noalign{\smallskip}
\hline

\noalign{\smallskip}
\CIIw{}&N(\CII{}) [$10^{17}$\,\cmsq]&5.25&5.18&3.00&1.47&2.30\\
&X(\CIIw{}) [$10^{-5}$]&2.5&0.43&0.81&0.057&0.48\\

\noalign{\smallskip} \hline \noalign{\smallskip}
&[C$^+$]:[C$^0$]:[CO] [\%]   &       26:16:58 & 10:12:78 & 16:12:72 & 03:10:87 & 05:09:85 \\
\noalign{\smallskip} \hline \noalign{\smallskip}
\end{tabular}
\end{center}
\end{table*}

\subsubsection{Carbon Monoxide}

The amount of $^{13}\RM{CO}$ is determined in two ways assuming again optically thin emission.
Firstly, by using $T_{\RM{ex},^{13}\RM{CO,high}}=79.4\,\RM{K}/\ln{(4/R_{^{13}\RM{CO,high}})}$
(Eq. (\ref{coratio})) with the $^{13}$CO 6\TO5 to 3\TO2 integrated intensity
line ratio $R_{^{13}\RM{CO,high}}$
and by Eq. (\ref{column13co}).
This ratio is a sensitive temperature measure from $\sim$20\,K up to a few 100\,K.
The excitation temperatures\HIDE{listed in Table \ref{ci-table}} vary between 30 and 46\,K and represent upper limits
because of possible self-absorption due to a high optical thickness in the lower transition.
The temperatures agree with values from \citet{wilson1990} who find $20-65$\,K from NH$_3$ emission,
and are always below the \CIw{} temperatures from above indicating an origin from different gas.
The $^{13}$CO column density, {\NEW $23-54\cdot 10^{15}$\,\percc{}, varies by a factor of 2 as the \CIw{} column density}.
The same scatter is found if we estimate the column density from $^{13}$CO 1\TO0 rather than $^{13}$CO 6\TO5 (Eq. (\ref{column13colow}))
and derive temperatures by Eq. (\ref{coratiolow}).
Those column densities tend to be higher, between $33-76\cdot 10^{15}$\,\percc{}, {\NEW indicating} that a larger amount of gas traced by
$^{13}$CO is cold.

Considering the optically thin emission of C$^{18}$O,
we estimate the total molecular gas column density from Eq. (\ref{columc18o})
using both $^{13}$CO T$_\RM{ex}$ estimates as limits.
We find high C$^{18}$O column densities of about $1-2\cdot 10^{16}$\,\cmsq{} along the ridge
and $2-4\cdot 10^{15}$\,\cmsq{} toward E and W.
\LIT{
\citet{garden1991b} determine a slightly higher average column density of $2.1\cdot 10^{16}$\,\cmsq{} toward the DR21 cloud core.
}

%

\subsubsection{Ionised Carbon}


In order to get a rough estimate of the C$^{+}$ column density from the ISO data, we
assumed $N(\CIIw{})=6.4\cdot 10^{20} \times I(\CII{})$\,[\cmsq] $($erg\,s$^{-1}$\,cm$^{-2}$\,sr$^{-1})^{-1}$
\citep{crawford1985}, which is a good approximation for
an excitation temperature of $>91$\,K, and a density $>n_\RM{cr}=5\cdot10^3$\,\percc{}.
The peak column density derived this way, $\RM{N}(\CIIw{})\sim 5.2\cdot 10^{17}$\,\percc{}, is {\NEW observed} towards DR21\,C{} and E{}, while
only about 28\,\% of the peak value is found towards DR21\,(OH).

\subsubsection{Abundances}
%
%

The comparison between C$^{18}$O and  $^{13}$CO column density (for the lower T$_\RM{ex}$ limit
in Table \ref{ci-table}) shows that the abundance ratio $[^{13}\RM{CO}]:[\RM{C}^{18}\RM{O}]=6.6-11.4$ 
is typically near the canonical local abundance ratio $[490]:[65]=7.5$ \citep{langer1990}.
\OUT{
With the exception of DR21\,E{}, where the ratio is slightly enhanced to 13.8,
this simple analysis already demonstrates that probably at least $\sim 18-26$\,\% of the detected column is warm.
A correction for an optical depth close to $\tau\sim1$ in $^{13}$CO would further enhance this
ratio and hence further increase this estimate.
{\MOD
This supports the scenario that parts of the molecular gas form a warm component while the rest is rather cold:
}
The cold gas is prominent with high column densities in the northern part and to the south along the ridge.
At the DR21 core and especially toward the east part, strong emission in submm- and far-IR CO and $^{13}$CO lines gives clear evidence
for the presence of warm molecular gas.
}

The atomic carbon abundances [\CIw{}]/[H$_2$] fall in the range of $0.1-1.5\cdot10^{-5}$.
{\NEW The high values} are only observed at the E and W positions.
In the cooler northern positions the abundance is reduced by a factor up to 10.
This is in contrast to the values of
$1-2 \cdot 10^{-5}$ by \citet{keene1997} found in cold dark clouds.

Except for DR21\,FIR1, there is a good correlation between X(\CIw{}) and X(C$^{18}$O).
Toward DR21\,(OH),
{\MOD the C$^{18}$O and \CI{} abundance are systematically lower than toward DR21 by factors $2-3$.}
Although, the position with the lowest \CI{} abundance
{\NEW ($10^{-6}$)}, FIR1, has the highest C$^{18}$O abundance 
{\NEW (up to $3.7\cdot 10^{-7}$)}.

\citet{mookerjea2006} present a compilation of the C/CO abundance ratios versus H$_2$ column densities for Galactic star forming regions and diffuse clouds.
The ratio at the positions DR21\,E{}, W{}, and FIR1, 0.07--0.17, agrees well with the relation
$\log[N(C)/N(CO)]=-0.94 N(H_2)+19.9$ given there for Cepheus B.
DR21 and DR21\,(OH) deviate from this relation but still are within the observed scatter.
We therefore
underline their conclusion that \CIw{} is no straightforward tracer of total H$_2$ column densities and total masses.

\OUT{
It is possible that we overestimated T$_\RM{ex}$.
The extreme case of 10\,K would reduce the $^{13}$CO column density by about 20\,\%.
}

{\NEW
The amount of carbon in the carbon bearing species C$^+$, C$^0$, and CO
can be compared using their relative abundances (last row of Table \ref{ci-table}).
The neutral atomic carbon relative abundance is rather homogeneous at about $9-16$\,\% of the total carbon gas content at all five positions}.
The rest is shared between CO and \CIIw{}.
Ionised carbon has the highest abundance
toward DR21\,E {\NEW (26\,\%)}.
Here, the gas is likely exposed to a high UV radiation field and carbon species are photodissociated to \CIw{} and ionised further to \CIIw{}.
High fractional CO abundances {\NEW ($>$85\,\%)}, such as for DR21\,(OH) and FIR1 were also reported for S\,106 \citep[$>$86\,\%, ][]{schneider2003} and W3\,Main \citep[$>$60\,\%, ][]{kramer2004}.
The ratios in other sources are, e.g., 37:07:56 for IC 63 \citep{jansen1996},
NGC2024 with 40:10:50 \citep{jaffe1995}, or
$\eta$Carina with 68:15:16 \citep{brooks2003}.

%

\subsection{Radiative transfer modeling}
\label{rad-model}

%

\begin{table*}[ht]
\caption[]{
The physical parameters for the observed ISO/LWS positions. 
The radii listed in column 2 give the inner and outer edge of each shell.
Column 3 to 7 lists the temperature, clump-averaged number density of H$_2$, volume filling, column density
of H$_2$, and the gas mass.
For these columns, the first entry is the mass-averaged value for the whole cloud, values in parentheses show averages within each shell.
The last column lists the reduced $\chi^2$ fit result for 11 lines (6 KOSMA, 2 FCRAO \& 3 ISO).
} \label{tab-carbon-content}

\begin{tabular}{lccccccc}
\noalign{\smallskip} \hline \hline \noalign{\smallskip}
  Position &Radii&T$_{\mathrm{gas}}$&n(H$_2$)&$\phi_v$&N(H$_2$) & Mass & $\chi_0^2$\\
 &[pc]&[K]&[$10^4$ \percc]&volume filling$^b$&[$10^{21}$\,\cmsq]$^a$ &[\msol (D/1.7\,kpc)$^2$]$^c$\\
\noalign{\smallskip} \hline \noalign{\smallskip}
DR\,21 E{} &0.04/0.2/0.3&62 (87/55)&92 (300/4.8)&0.04 (0.01/0.89)&46 (9/37)& 243 (46/197) & 2.2 \\  

DR\,21 C{} & 0.12/0.2/0.6 & 36 (148/32) & 110 (40/110)& 0.024 (0.08/0.02) &65 (2/63) & 1379 (51/1328) & 3.5\\ 



DR\,21 W{} &0.04/0.1/0.3&54 (118/52)&12 (100/9)&0.59 (0.06/0.8)&88 (3/85)&489 (15/474) & 2.8 \\ 

DR\,21 (OH)&0.04/0.2/0.7& 31 (99/29)& 161 (220/160)&  0.01 (0.01/0.01)&49 (1/48)&1421 (34/1387) & 3.9\\ 

DR\,21 FIR1&0.04/0.2/0.5&41 (82/37)&16 (220/2)& 0.1 (0.01/1)&47 (4/43)&700 (59/641) & 4.1 \\ 

\noalign{\smallskip} \hline \noalign{\smallskip}
\end{tabular}
\begin{list}{}{}
\item[$\NOTE{a}$] Column densities are source averaged and not normalized to $80''$.
\item[$\NOTE{b}$] The product of the volume filling and the abundance of the species is one of the free parameters. We fixed the abundance
[X/H$_2$] of $^{12}$CO to $2\cdot 10^{-4}$ (resp. $3.1 \cdot 10^{-6}$ for $^{13}$CO, and $4 \cdot 10^{-7}$ for C$^{18}$O).
\item[$\NOTE{c}$]
The molecular gas mass is derived from the cloud-averaged density.
This density is the product of local H$_2$ density of the clumps and volume filling correction $\phi_v$.
\end{list}
\end{table*}

%


In a self-consistent approach, the
temperature, column and number density, total molecular mass, volume filling
and mass composition were determined using a 1-D radiative transfer code
\citep[{\sc SimLine}, ][]{ossenkopf2001}.
In order to describe these physical conditions,
we created heterogeneous models with
internal clumping and turbulence in a spherically symmetric geometry for each of the 5 observed regions.
The physical parameters of a shell are defined by their values at the inner radius and the radial dependence is described by a power-law.

A model with two nested shells provided in most cases reasonable results:
{\it (i)} a hot and dense molecular {\em interface zone} wrapped around a central \HII{}-region,
{\it (ii)} enclosed by a massive cold cloud -- the {\em envelope region}.
While in the inner component primarily the mid- and high-$J$ CO emission lines are excited,
the envelope is responsible for self-absorption and the low lying lines of $^{13}$CO and C$^{18}$O.
We set the abundance of species with respect to molecular hydrogen to standard values (see Table \ref{tab-carbon-content}).
The source size is a crucial parameter that is not well confined by the line profiles or
the map we obtained.
A small cloud model can easily produce results similar to those for a bigger cloud with a lower volume filling.
We therefore approximated the radii using high resolution line and continuum observations
\citep[][]{wilson1990,lane1990,jaffe1989,chandler1993,vallee2006}.
The inner cut-off was assumed to be 0.04\,pc ($<10^4$\,AU), except for
DR21\,C where we estimated $\sim$0.12\,pc from the 14.7\,GHz continuum \citep{roelfsema1989}.
The \HII{}-region is considered to be free of molecular gas and we neglect the continuum contribution by ionised gas.
Gradients within the shells in any of the parameters are initially set to zero
and are adjusted during the fit if required.

We used the following {\em recipe} to find appropriate model parameters:
{\it (i)} The linewidth is approximated by adjusting the turbulent velocity field until both the low-$J$ and mid-$J$ line wings match the observed profile.
{\NEW As line emission from the inner warm gas is shielded by the colder envelope gas closer to the surface,
the width of the CO absorption dip in the line center constrains the turbulent velocity dispersion of this envelope component.}
{\it (ii)} The H$_2$-density is defined as a local clump density
parameter and varied between $10^4$ and $10^7$\,\percc{} independently for both shells.
The high densities are a pre-requisite for high-$J$ CO excitation.
{\it (iii)} 
{\NEW
In the optical thin line wings we see predominantly material from the inner shell where the
simulated macroturbulent motion leads to
the higher line width
which is only partly shielded by the outer shell.
This allows us to fit the temperatures of both the inner and the outer shell from the line shape.}
{\it (iv)} The model allows to differenciate between radial and turbulent gas motion.
Where lines of high opacity toward the line center show a blue-skewed profile,
we introduced radial infall as a plausible explanation.
If possible, we tried to keep the radial component close to zero.

The full line profiles of CO, $^{13}$CO, and C$^{18}$O from KOSMA and FCRAO are taken into account in a
reduced $\chi^2$ fit in $\sim$1900 degrees of freedom and using $20-28$ free parameters.
The second velocity component is masked out by a window centered on $9 \pm 3\,\kms$.
For the ISO lines, we fit the integrated flux of three lines.
CO 14\TO13 is not considered because no consistent fit was possible in conjunction with the other CO lines.
In total, we include 11 lines weighted by the $\sigma^2$ determined
{\NEW from a baseline region not including any lines.
The spectra do not contain any obvious baseline ripples above the system noise.
}

\subsubsection{Line Modelling: CO, $^{13}$CO, and C$^{18}$O}

{\NEW
We independently optimized the model parameters for each position.
Fig. \ref{plot-spectra} shows the derived line profiles.
%
At all five positions, we obtain solutions of better than $\chi_0^2=4.1$ which we
consider -- given the number of fitted lines -- a very good result, and definitely
much better than for a single-component model.

Within the interface-zone,
a temperature range from 80\,K (at DR21\,FIR1) to 148\,K (at DR21\,C) for gas masses of $15-60$\,\msol{} is found.
This temperature is much higher than the typical LTE excitation temperature of the mid-$J$ CO line peak.
The fit converges toward solutions with a high density of about $10^6$\,\percc{} and
with a volume filling of a few percent, indicating a non-homogeneous distribution.
The warm gas dominates the emission in the line wings and therefore is explained by
the steep turbulent velocity gradient with velocities up to $30\,\kms$.
Because the envelope is much colder and more massive (cf. Sect. \ref{massfilling}),
the average temperature in the regions
DR21\,C{} with 36\,K and DR21\,(OH) with 31\,K is rather low compared to the E, W, and
FIR1 position. There, the gas temperature rises up to 60\,K but the gas masses are only a few hundred
solar masses.
Besides the mass fraction,
the steepness of the gradient in excitation condition and line width
between inner and outer component determines whether we
see a distinct self-absorption (e.g., DR21\,C) or a flat-top (e.g., DR21\,W) line profile.
Table \ref{tab-carbon-content} compiles a list of the derived physical parameters.
A detailed discussion of the individual positions is given in App. \ref{rad-model_pos}.
}


\subsubsection{Line Modelling: \CI{} and CS}
\label{ciandcs}

As a consistency check, we computed the CS and \CIw{} profiles using the same model parameters as derived from CO.
This implies a homogeneous distribution of CS and \CIw{} with CO, only affected by the abundances of the species.

For both species we {\NEW also} obtained excellent fit results.
The resulting line profiles are shown in Fig. \ref{plot-spectra}.
CS 7\TO6 is only excited where the H$_2$ density and temperatures are sufficiently high.
Hence, we find CS emission in the dense regions DR21\,C{} and (OH) only.
The models fit best to the observations for an abundance of X(CS)=$4\cdot 10^{-9}$ \citep{hatchell1998},
whereas the range
$1\cdot 10^{-10}$ to $2.3\cdot 10^{-10}$
given by \citet{shirley2003} is clearly too low for this region.

Instead of a carbon abundance variation across the positions (as found in Sect. \ref{civsco}),
an abundance X(\CIw{}) of $1\cdot 10^{-5}$ nicely fits our \CI{} observations.
The \CIw{} excitation temperatures {\NEW as derived in Sect. \ref{atomiccarbon}} 
fall in the range covered by the two-component radiative transfer model.

%

\begin{figure*}[th]
\includegraphics[width=0.221\linewidth,angle=-90]{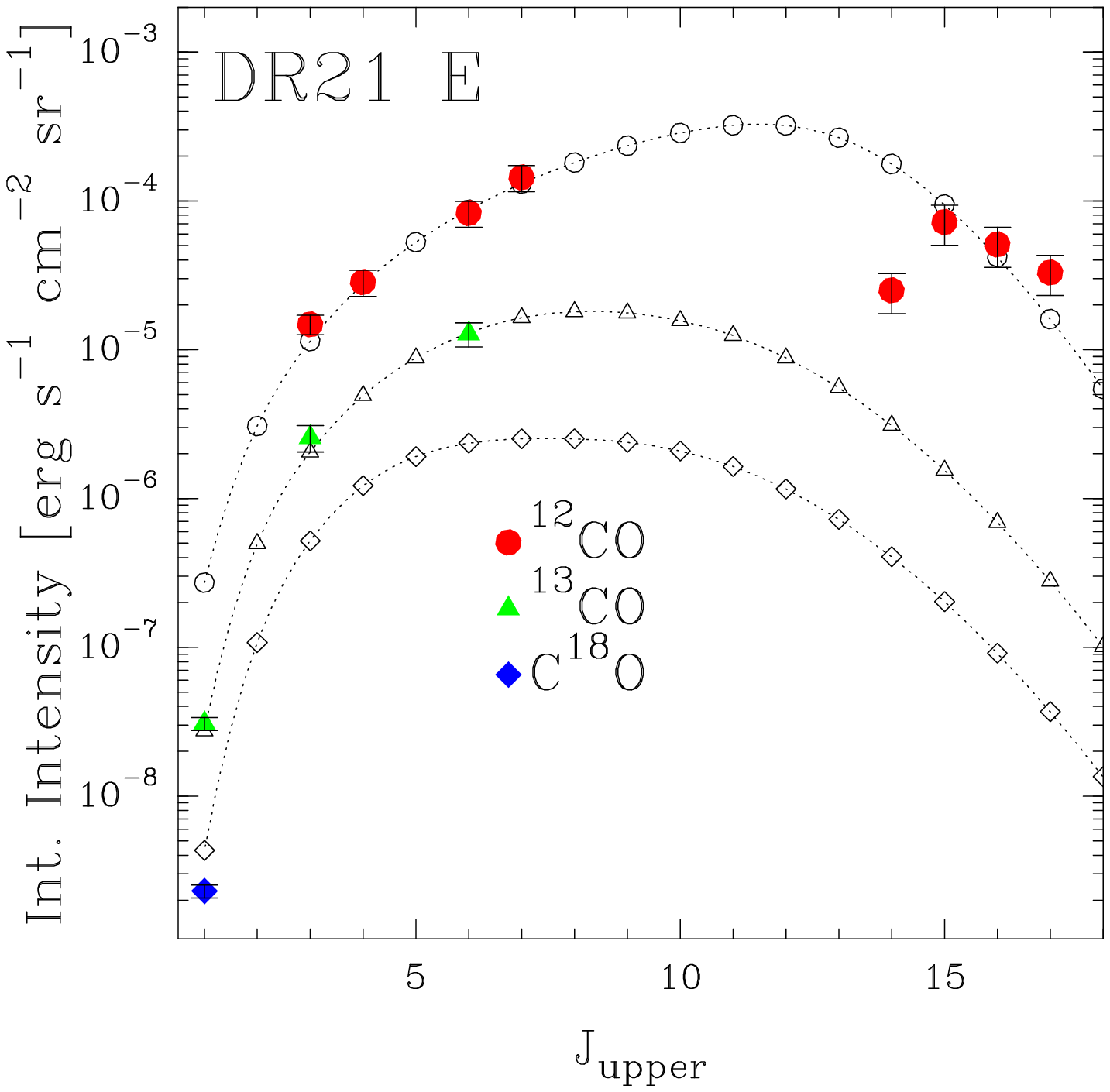}\nobreak
\includegraphics[width=0.221\linewidth,angle=-90]{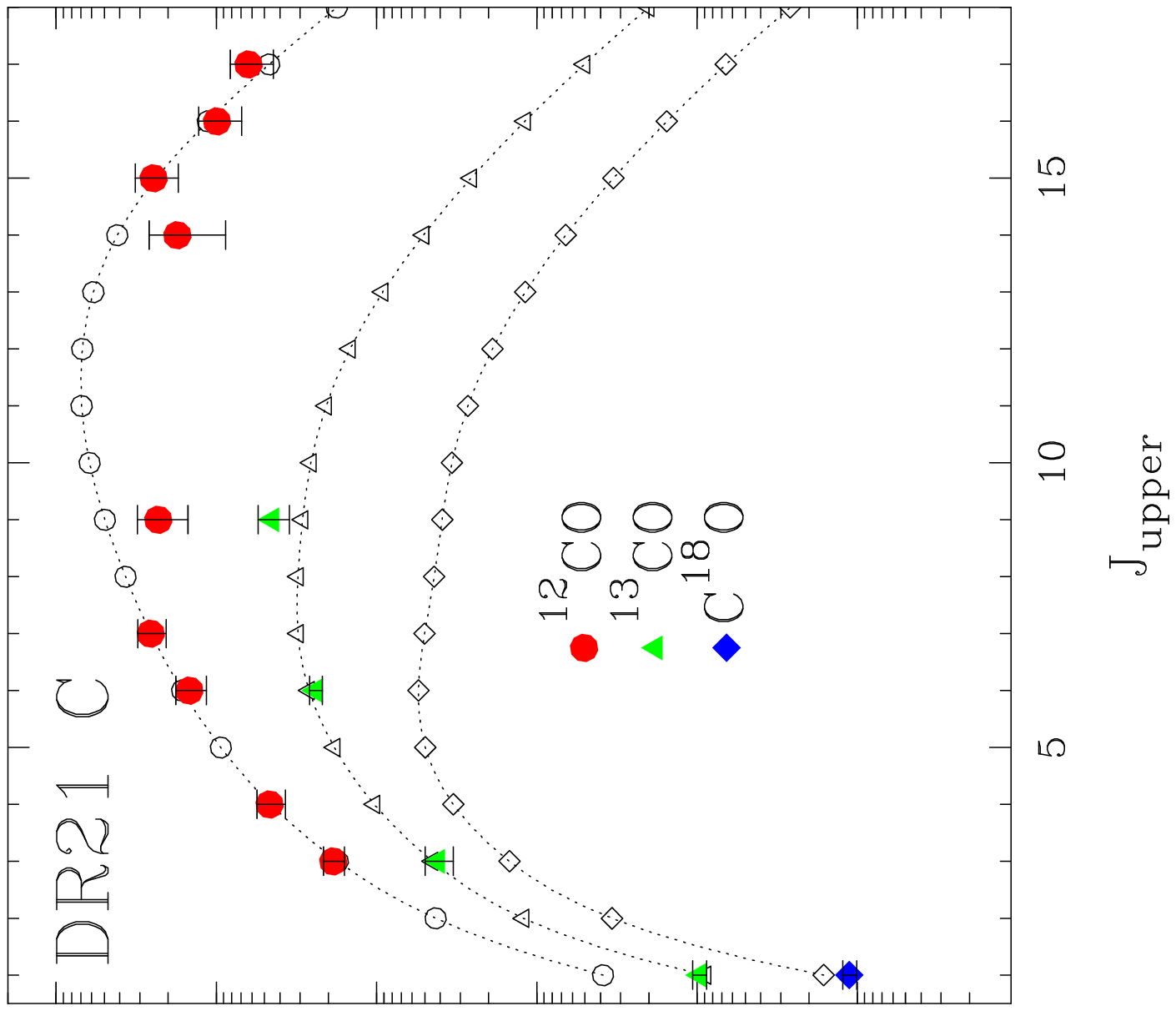}\nobreak
\includegraphics[width=0.221\linewidth,angle=-90]{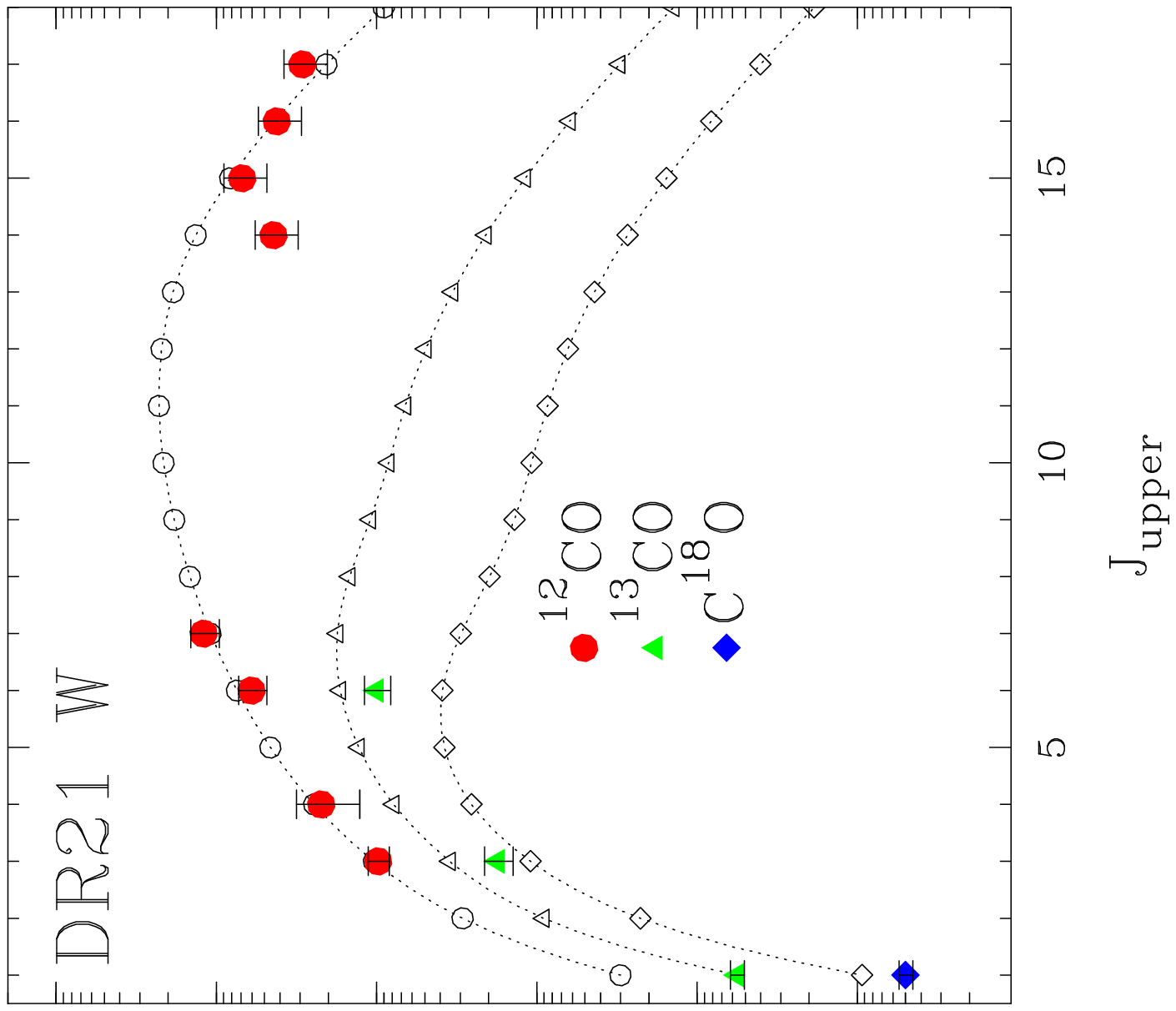}\nobreak
\includegraphics[width=0.221\linewidth,angle=-90]{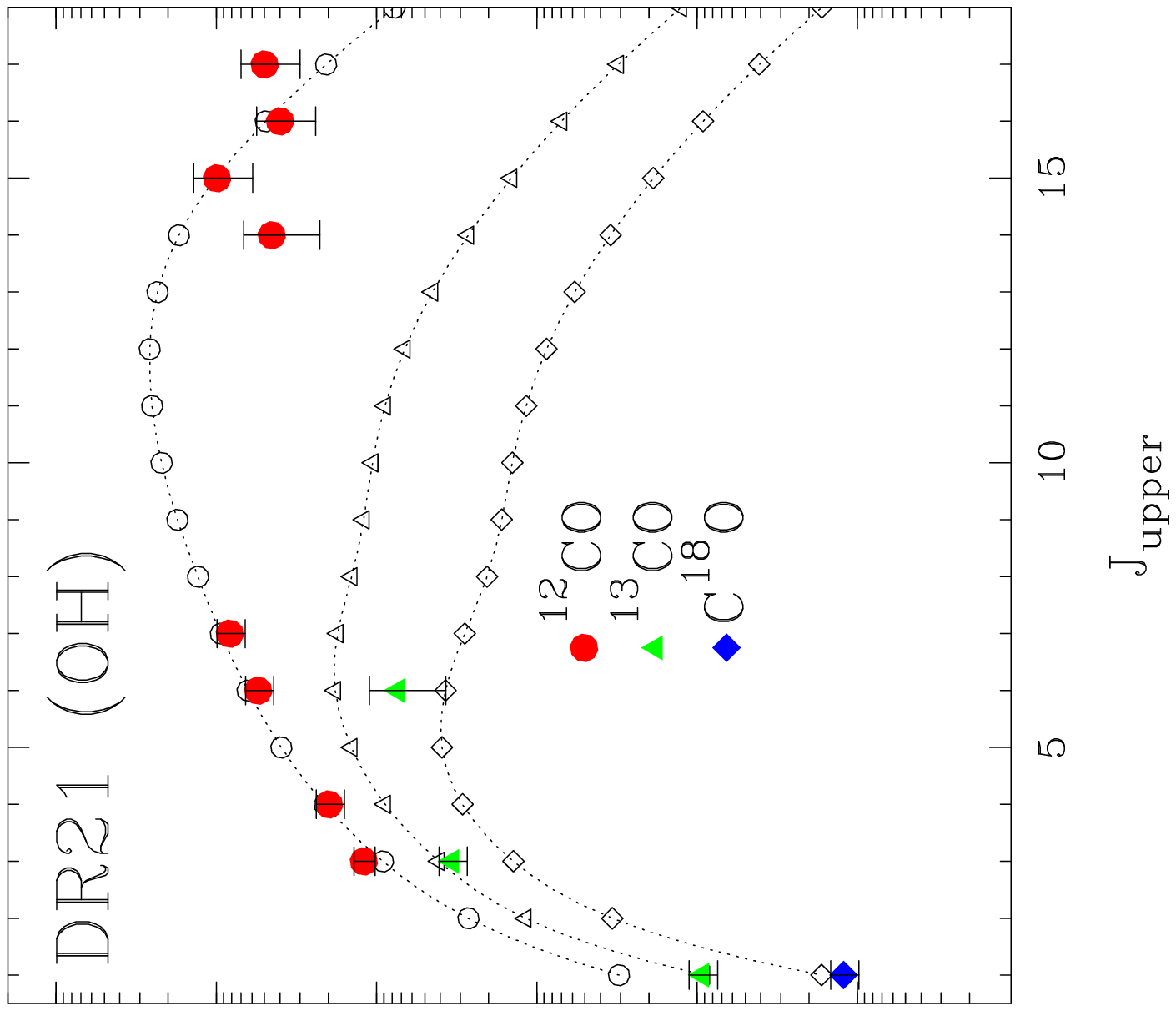}\nobreak
\includegraphics[width=0.221\linewidth,angle=-90]{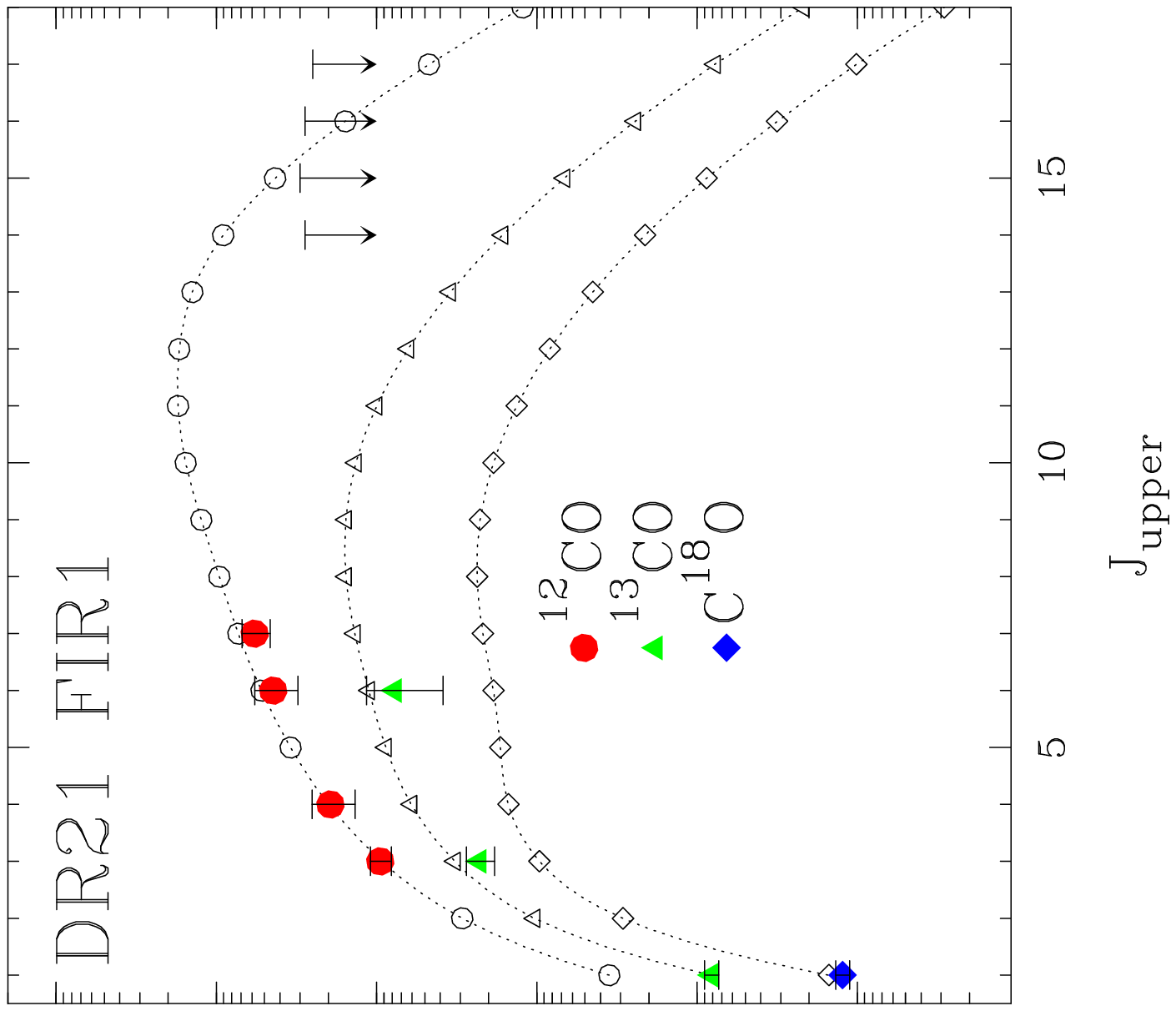}\nobreak
\caption{
Cooling curves of the best-fit model line fluxes (marked by open symbols) of CO, $^{13}$CO, and C$^{18}$O.
Observed line intensities are indicated as filled symbols (see Table~\ref{tab-pdr-lines}).
At position DR21\,C{}, we included two observations by \citet{boreiko1991} of CO and $^{13}$CO  $J=9\TO8$.
Since the  $^{13}$CO 6\TO5 map does not cover the postitions DR21\,(OH) and DR21\,FIR1 on a fully sampled grid,
we expect a larger error. 
The high-$J$ CO emission at DR21\,FIR1 are upper limits.
}
\label{fig_cooling}
\end{figure*}

\subsubsection{Masses and Volume Filling}
\label{massfilling}

The volume filling, $\phi_v$, is a model parameter capable of increasing the clump density
without increasing the overall mass as well.
The determined $\phi_v$ is usually low between 1\,\% to 8\,\% (Col. 5 in Table \ref{tab-carbon-content}).
At DR21\,C, this is confirmed by our observations of the CS 7\TO6 transition
which is sensitive to densities of several $10^6$\,\percc{}:
Given the high H$_2$ column density ({\NEW up to $2\cdot 10^{23}$\,\cmsq{})} 
this translates to a column-to-volume density ratio of about 0.006\,pc.
If we assume that the visible extent of the DR21 source ($\sim$$0.69$\,pc in CS, after deconvolving the beam) is similar
to the size along the line-of-sight, a volume filling of $\sim$$0.01$ is derived.
A high degree of small scale spatial and velocity structure is also needed to explain the hyper-fine-structure
anomalies observed in ammonia data toward DR21 \citep{stutzki1985}.
\citet{shirley2003}, who studied CS 5\TO4 in a sample of high-mass star-forming regions,
derived a volume filling with a median of $0.14$.
This is one order of magnitude higher than our results for the dense regions DR21 and DR21\,(OH).

The total mass for all 5 positions is $4.2\cdot 10^3$\,\msol{}, and mostly shared between DR21\,C and DR21\,(OH) with $\sim$$1.4\cdot 10^3$\,\msol{} each.
In direct comparison to the masses derived from dust observation in Sect. \ref{greybodyfit}, we find that our model
gives higher masses
($\times1.6-\times1.8$), except for the DR21\,(OH) model mass.
To compare our mass estimate with masss from the literature, often derived
for a different distance, we have scaled the mass in Table 
\ref{tab-carbon-content} with the square of the distance, basically
assuming that the column density, computed from the observed lines, is
independent from the distance. This approach is consistent with a 
local radiative transfer treatment, such as LTE or escape probability
analyses, where the molecular excitation does not depend on the total
gas mass. For the self-consistent radiative transfer code, this
assumption is, however, not completely justified, but deviations
from this simple scaling relation are to be expected. When comparing
the values from Table \ref{tab-carbon-content} to results obtained for
much larger distances our quadratically scaled mass estimates are
therefore always expected to be too high and when comparing them
to results for lower distances they should be systematically lower.
\citeauthor{shirley2003}
estimated virial masses of 561\,\msol(D/1.7\,kpc) for DR21 (resp. 714\,\msol(D/1.7\,kpc) for DR21\,(OH)).
These estimates are lower than our masses
but were calculated over a much smaller radius of 0.17\,pc(D/1.7\,kpc) while our region covers up to 0.7\,pc.
\citet{ossenkopf2001} estimated the DR21 mass using a radiative transfer analysis on CS data.
They estimated 578\,\msol(D/1.7\,kpc)$^2$ at a (cloud-averaged) density of $1.6\times 10^3$\,\percc{}.
Their lower density is partly explained by the larger outer radius,
that was chosen to be 1.0\,pc(D/1.7\,kpc).
It is probably also due to the larger physical cloud size at the 
assumed distance of $D=3$\,kpc which prevents the simple quadratic
distance scaling of the mass in a self-consistent radiative-transfer
treatment.

%
The total mass of DR21\,(OH)
corresponds to a cloud-averaged column density of $\sim$$5\cdot 10^{22}$\,\cmsq{}.
The value of $5.7\cdot 10^{23}$\,\cmsq{} reported by \citeauthor{richardson1994} from 1100\,\micron{} data
and our value given in Table~\ref{iso-dust} based on the far-IR dust-emission are higher by a factor of $5-10$.
A flatter spectral index $\beta$ may explain why dust and line observation analyses result in different conclusions.
Values between 1.0 \citep{vallee2006} and 2.0  \citep{chandler1993} were discussed for DR21\,(OH) in the literature.
If $\beta$ is lower than the assumed 1.5, this will have a significant effect on the derived $\RM{A}_v$.
For instance, with $\beta=1.25$, the column density is reduced to about one fourth of our value at $\beta=1.5$,
then is consistent with the CO model results.


An alternative explanation for the discrepancy could be depletion
as a result of freeze-out of CO onto grains.
Some authors \citep[e.g., ][ and references therein]{vallee2006} use a depletion factor of 10 or more (X(CO)=$2.5\cdot 10^{-6}$ at DR21\,(OH))
as appropriate for star-forming clouds.
In \citet{jorgensen2004}, the abundance is reduced by a factor of 10 for the conditions $T<30$\,K and $n>3\cdot 10^4$\,\cmsq{}.
The DR21\,(OH) envelope and the resulting column density would comply with these conditions.
However, the CO line data analysis does not allow us to draw conclusions toward depletion.

%
\HIDE{
Fractions larger than 1 are found in some of the outer shells.
Explanation may be an enhanced CO abundance or is due to the geometrical nature of the real cloud where additional material outside the spherical limits is needed.
}

\OUT{
The pressure, expressed as $P/k_\RM{B}=n(\RM{H}_2)\cdot T$ for the cloud-averaged values,
is typically $>4\cdot 10^7$\,K\,\percc{}.
At DR21\,W{} and toward FIR1 the pressure drops to $\sim$$6\cdot 10^6$\,K\,\percc{}.
The interface regions show even higher pressures up to $10^8$\,K\,\percc{},
a value that fits well to the median pressure determined by \citeauthor{shirley2003} of $1.5\cdot 10^8$\,K\,\percc{}.
}

\subsubsection{Problems of the fit}

As noted before,
a precise determination of the outer radii of the model clouds, and hence their mass,
is quite uncertain and a small increase may already more than double the mass. In the models of DR21\,C{} and DR21\,(OH),
we found no reasonable solutions for radii much smaller than 0.6\,pc matching the $^{13}$CO and C$^{18}$O 1\TO0 observations.
%
{\NEW
As possible indication of a systematic error in the column density
we typically see that
the observed peak temperature of the C$^{18}$O 1\TO0 line is weaker than predicted
by the models (see Fig. \ref{plot-spectra}) while the $^{13}$CO 1\TO0 line is well reproduced.}
This can either be a result of an inaccurate modeling of density and temperature conditions,
or our assumption about the fixed abundances for CO, $^{13}$CO, or C$^{18}$O
is incorrect.

The models overestimate the $^{13}$CO 3\TO2 peak temperature.
Except for DR21\,E{}, this transition is optically thick ($\tau>2$) and reflects the kinetic temperature of the envelope gas.
As the CO 3\TO2 to 7\TO6 lines also become optically thick in the envelope,
the brightness temperature at the self-absorbing line center of these mid-$J$ lines is governed by
the same material.
The very deep ($\sim$5\,K) absorption dip in CO 7\TO6 observed by \citet{jaffe1989} indicates very cold foreground gas ($<20$\,K)
which we cannot reproduce with only one component for the outer shell.
A better match of the observed intensities may be possible by adding a third colder layer outside of the envelope.




Since the cold and warm components certainly are intermixed to some extent rather than separated in two nested shells,
the focus during the modeling was put on a qualitative description of the different components and should not be seen as a
geometric representation of the real cloud structure.
We demonstrated, however, that the simplified assumption of a spherical cloud is adequate to estimate physical properties, such as mass, density,
filling factors, and temperature.
\OUT{
In context of PDR or stock heated emission, it may be very helpful to weight the contribution from a cold and a warm components consistently.
}

%

\subsection{Oxygen and Nitrogen}
\subsubsection{ISO/LWS far-IR lines: \OI{}, \OIII{}, and \NII{}}

\OUT{
The LWS band covers many important far-IR cooling lines.
These are at the same time important tracers of Photon Dominated Regions.
}
In addition to the high-$J$ CO lines (see Sect. \ref{isocolines}), ISO provided the integrated line intensities of
\OI{} (63 and 145\,\micron{}) (Table \ref{tab-iso-pdr-lines}).
Their relative strength and cooling efficiency depends on the temperature and density regime:
At low temperatures, rotational mm- and submm-lines of CO dominate the cooling of the molecular and neutral gas.
At temperatures above T$\sim 150$\,K, pure rotational transitions of OH, H$_2$O, H$_2$, and the 63\,\micron{} \OI{} line become more important.
\OI{} at 63\,\micron{} and the atomic and ionic fine-structure lines of \CII{} and \CI{} are
diagnostic lines for PDRs.
Due to the nearby \HII{}-regions, the ionic lines \OIII{} and \NII{} are also expected to be strong emitters and may serve as diagnostics
to discriminate the fraction of \CII{} emission coming from the neutral and ionised gas.
For completeness, we included the \OIII{} (52 and 88\,\micron{}) and \NII{} (122\,\micron{}) lines in the table.



Since DR21\,E{}, C{} \& W{} line up along the outflow axis, we compare the emission in the outflow
lobes with the central position:
The strong \OI{} emission at $63$\,\micron{} decreases rapidly toward the west \citep[cf.,][]{lane1990}.
This line easily becomes optically thick and \citet{poglitsch1996} indeed found a deeply absorbed line toward DR21.
The 63\,\micron{} emission extends toward the E{} position at almost the same strength as toward the center.
As the A$_\RM{v}$ at the same time drops down, the gas traced by \OIw{} must be a separate, necessarily warm component.
\OI{} at $145$\,\micron{} is peaked at DR21\,C{}, showing only a slight east-west asymmetry.

If we assume thermalized optically thin \OI{} emission, we can derive the excitation temperature from
T$_\RM{ex}=227.7\,\RM{K}/\ln(2.83/(12.22\cdot R_\RM{\OI{}}))$
(Eq. (\ref{oiratio})), where
$R_\RM{\OI{}}$ is the ratio of \OI{} 145/63\,\micron{} emission (cf. Table \ref{tab-iso-pdr-lines}).
The ratios of observed intensities, 0.06--0.12, result in temperatures between 170 to 370\,K and are to be seen
as strict upper limits since $R_\RM{\OI{}}$ decreases for optically thick \OI{} 63\,\micron{}
line emission.
The \OI{} line ratio at DR21\,(OH) of $R_\RM{\OI{}}$=0.29 requires a temperature just below 100\,K.
We detect the weakest emission of \OI{} at DR21\,FIR1.
Errors in $R_\RM{\OI{}}$ become too large to estimate
a temperature here.

As expected for the position of the \HII{}-region,
the ionised \OIII{} lines both are strong at DR21\,C{} and decreases to the sides.
Furthermore, the \OIII{} 88/52\,\micron{} line ratio is a good discriminant of the 
electron density in the range $10^2-10^5$\,\percc{}, while being insensitive to abundance and temperature variations \citep{rubin1994}.
At DR21\,C{}, a ratio of about 0.4 indicates an electron density of $\langle N_e\rangle \sim{}10^{3}$\,\percc{}.
In the east, the ratio rises to 0.8 indicating an extension of the \HII{}-region to this side.
To the west, the ratio is above 1.4.
An electron density below $10^2$\,\percc{}, consistent with the observed \OIII{} line ratio at DR21\,(OH),
indicates a low degree of ionisation.
No \HII{}-region has been identified in this area, e.g., in the VLA 6-cm continuum
\citep{palmer2004}. Thus, the finding of \OIII{} emission is a bit surprising.
The \OIII{} ratio $>3\pm 1.3$ found at DR21\,FIR1 is not covered by the \citeauthor{rubin1994} model but corresponds to a very low $\langle N_e\rangle$.

\NII{} 122\,\micron{} was detected at the E{} and W{} positions only. At the positon of the \HII{}-region DR21\,C{},
the $3\sigma$ noise limit still permits centrally peaked emission.
The fraction of \CII{} stemming from the \HII{}-region can be estimated from the intensity ratio of \CII{} to \NII{}.
The Galactic abundance ratio C/N in dense \HII{}-regions
leads to an expected ratio of \CII{}/\NII{}=1.1 \citep[][ and references therein]{kramer2005}.
With the observed \CII{}/\NII{} ratio between 11--19 for DR21, less than 10\,\% of the \CII{} emission are estimated to come from the ionised gas.

%

\begin{table*}[ht]
\caption[]{
Total cooling intensities of CO, \CI{}, \CII{}, and \OI{}. For CO we use the
best fitting model (Sec. \ref{rad-model}) and sum over all rotational lines up to
$J$=$20$.  For \CIw{} and \OIw{}, we sum the observed intensities of both fine
structure lines respectively.
The values in parentheses show the relative
contribution to the sum $I_\RM{tot}$.  $I_\RM{IR}$ is the total
observed infrared intensity (Table \ref{iso-dust}).  The total
cooling efficiency $\epsilon_\RM{tot}$ is defined by
$I_\RM{tot}/I_\RM{IR}$.  For comparison, we also list the $\epsilon'$
defined by $(I_\RM{\CII}+I_\RM{\OI})/I_\RM{IR}$.
}\label{cooling-table}
\begin{center}
\begin{tabular}{lcccccc}
\noalign{\smallskip} \hline \hline \noalign{\smallskip}
Intensity in [$10^{-4}$ erg\,s$^{-1}$\,cm$^{-2}$\,sr$^{-1}$]&DR21\,E{} & DR21\,C{} & DR21\,W{} & DR21\,(OH) & DR21\,FIR1\\
\noalign{\smallskip} \hline \noalign{\smallskip}

\noalign{\smallskip}
%
$I_\RM{CO}(=\Sigma$ CO$_{1\le J_\RM{u}\le 20}$)
  & 24.1 (28.8\,\%)&  51.3 (44.4\,\%)& 18.7   (53.7\,\%)&20.0  (75.9\,\%)& 13.4  (68.7\,\%)\\
$I_{\CI}(=\Sigma$ \CI)                
  & 0.22 (0.3\,\%) &  0.45 (0.4\,\%) & 0.23   (0.7\,\%) & 0.25 (0.9\,\%) & 0.27   (1.4\,\%)\\
$I_{\OI{}}(=\Sigma$ \OI{})              
  & 51.1 (61.1\,\%)&  55.6 (48.2\,\%)& 11.2   (32.1\,\%)& 3.78 (14.4\,\%)& 2.23   (11.4\,\%)\\
$I_{\CII{}}$                      
  & 8.2  (9.8\,\%)&   8.1  (7.0\,\%) & 4.7   (13.5\,\%) & 2.3  (8.8\,\%) & 3.6   (18.5\,\%)\\
\noalign{\smallskip}
$I_\RM{tot}$           
  & 83.62 (100\,\%)& 115.45 (100\,\%)& 34.83  (100\,\%) & 26.33 (100\,\%)& $<19.5$ (100\,\%) \\
$I_\RM{IR}$            
  & 5500            &33900            &1810              &12700           &3680\\
\noalign{\smallskip} \hline \noalign{\smallskip}
$\epsilon_\RM{tot}$ [\%]    &  1.5           &  0.34           & 1.9              & 0.21           & 0.52 \\
\noalign{\smallskip} \hline \noalign{\smallskip}
$I_{\CII+\OI}$ (\CII$+$\OI{}$_{63\,\micron{}}$)       & 56.4 & 57.5 & 14.8 & 5.24& 5.38\\
$\epsilon'$ [\%]             &  1.0 &  0.17 &   0.82 & 0.04 & 0.15\\
\noalign{\smallskip} \hline \noalign{\smallskip}
%
\end{tabular}
\end{center}
\end{table*}


\subsubsection{Line Modelling: \OI{} and \CII{}}

A determination of the \OIw{} and \CIIw{} abundances may be difficult.
If we assume that their abundance is closely connected to the dense molecular gas,
we can calculate and fit the emission of these two species
by adjusting the abundance of the two-shell radiative transfer models (Sect. \ref{rad-model}).
At DR21\,C{}, we find abundances of $6\cdot 10^{-5}$ for \OIw{} and $2.5\cdot 10^{-4}$ for \CIIw{}.
For both \CIIw{} and the \OIw{} 63\,\micron{} line, the opacity toward the line center is about 9 and the line profile is deeply self-absorbed.
The high energy above the ground level of \OIw{} leads to almost complete absorption of the center emission leaving signatures of warm gas only in the wings.
The 145\,\micron{} transition ($\tau \sim 0.5$) is unaffected and is a better measure of the abundance.
As \CIIw{} has only one transition and also shows some absorption in the model profiles,
the abundance may be less accurate.
At DR21\,(OH), the abundance of \OIw{} is $\sim$$1.2\cdot 10^{-5}$ and $9\cdot 10^{-5}$ for \CIIw{}.
The abundances at all five observed positions vary between $1-6\cdot 10^{-5}$ for \OIw{}, and $1-4\cdot 10^{-4}$ for \CIIw{}.

Compared to our LTE estimate in Sect. \ref{civsco}, the \CIIw{} amount derived here is higher by factors of $13-160$.
Because of recombination reactions, it is rather unlikely to have a large amount of \CIIw{} {\it hidden} in the cold gas, this can be considered as an
upper limit.

\OUT{
This would dramatically shift the fraction of \CIIw{} gas with respect to CO and \CIw{}.
In fact, the [C$^+$]:[C$^0$]:[CO] range, based on the model abundances, is 30:03:67 (in percent) at DR21\,(OH) to 65:02:33 at DR21\,E{}.
We see the reason for this high abundance in the assumption we made in Sect. \ref{civsco},
where we contributed {\em all} \CII{} emission to warm gas only.
Because of the high excitation temperature, a large amount of C$^+$ is {\em hidden} in the cold envelope gas and
just not visible in \CII{} emission.
However, C$^+$ is not stable under these cold dense conditions and should recombine to atomic carbon.
This clearly shows the limit of the here shown analysis and where chemical modeling is needed.
}

%
%

%


%

\section{Line cooling efficiency}
\label{enerybalance}
\label{line-cooling-eff} 

The relative importance of the various line tracers among each other
and with respect to the FIR continuum provides a detailed signature of
the heating and cooling processes in the DR21 molecular cloud.
Table \ref{cooling-table} lists the cooling intensities at the five
selected ISO positions E, C, W, OH, and FIR1.

The total line cooling intensity, $I_{\rm tot}$, derived by summing
the contributions of CO, \CIw{}, \CIIw{}, and \OIw{}, peaks at position C{},
dropping to less than 15\% of this peak value at FIR1.  

The relative contributions to $I_{\rm tot}$ vary strongly at the 5
positions. \OI{} is the strongest coolant at positions E{} and C{},
while CO is strongest at W{}, (OH), and FIR1, tying up more than
50\,\% of the total cooling. The relative contribution of CO to
$I_{\rm tot}$ varies between 26\,\% at E{} and 76\,\% at (OH) while
the relative contribution of \CII{} is lower, varying between 7\,\% at
C{} and 21\,\% at FIR1.  The two \CI{} fine structure lines contribute
generally less than 2\,\%.  The ratio of CO cooling relative to \CI{}
cooling, peaks at C{} at 114 and drops to less than 50 at FIR1.

We also compare $I_{\rm tot}$ with the FIR continuum to derive the
gas cooling efficiency $\epsilon_\RM{tot}$ as a measure of the fraction of
FUV photon energy which goes into gas heating.
$\epsilon_\RM{tot}$ varies by one order of magnitude,
between 0.2\,\% at (OH) and $1.5-2$\,\% at the two outflow positions
E{} and W{}.  

Table \ref{cooling-table} also lists the cooling efficiency
$\epsilon'$ just taking into account the 158\,\micron{} \CII{} and the
63\,\micron{} \OI{} lines, ignoring CO, as is often done in the
literature.  $\epsilon'$ varies between 0.04\,\% at (OH) and 1\,\% at
E.

Several studies in the literature have determined the gas heating
efficiency $\epsilon'$ in Galactic star forming regions and in
external galaxies.  \citet{vastel2001} find values of typically
$0.2-0.4$\% in Galactic star forming regions, $1.1$\,\% at the Orion
Bar, and less than $0.035$\,\% toward W49N. For W3\,Main/IRS5,
\citet{kramer2004} report $\epsilon'$ of $\sim 0.1$\,\% similar to
the value found in DR21\,C{}.  \citet{malhotra2001} find efficiencies
between $0.05$ and $0.3$\,\% in 60 unresolved normal galaxies.
\citet{kramer2005} find efficiencies between 0.21 and 0.36\,\% in the
centers and spiral arm positions of M83 and M51.

PDR models, e.g., by \citet{kaufman1999}, predict the heating
efficiency $\epsilon'$ of a cloud illuminated from all sides between
0.01\,\% to 1.5\,\% and our values are within these theoretical
limits.

A remarkable  point is that only a few percent of the mass (up to 20\,\% at DR21\,E{}) is efficiently heated.
{\NEW Considering only the cold envelope component of our radiative transfer models, i.e., removing the warm interface zone,
shows that the heated gas accounts for most of the CO emission
($80$\,\% to $94$\,\%, summing over $J_\RM{up}=1...20$) in terms of radiated energy.}

Since the total line power of all lines that we consider, $I_{\rm
  tot}$, does not include the contributions from other cooling lines
(e.g., of H$_2$, H$_2$O, OH, and PAHs) we may miss some of the cooling energy.



%

\section{Summary}

\OUT{
The observed region is known to harbour high mass star forming regions at different evolutionary stages (references?), and the maps confirm the diversity of this region.
The combination of different observations into a unique dataset allowed a deep view into this high-mass star-forming region.
}

We presented the first comprehensive mapping of sub-mm lines in the whole DR21/DR21\,(OH) region.
The line emission of \CI{} at 492\,GHz and 809\,GHz, CO 3\TO2, 4\TO3, 6\TO5, and 7\TO6, $^{13}$CO 3\TO2
and 6\TO5, and of CS 7\TO6
reveals contributions from different gas components:
{\it (i) } A very wide winged emission in mid-$J$ CO transitions having a line width of more than $15\,\kms$ and self-absorption, and
{\it (ii)} narrow (4 to $6\,\kms$) emission in the other lines.
Based on these lines, and supported by mm- and FIR-lines from FCRAO and ISO, we confined temperature, density, masses, and volume filling
for five positions toward DR21, DR21\,(OH) and DR21\,FIR1 using a radiative transfer code by modeling the
spectral line profiles.



1. {\NEW We found that 2-component models consisting of a warm inner region and a cold envelope} are needed to explain the
observed strengths of the energetically low and high lying transitions of CO and to explain the self-absorption.

2. The bulk of molecular gas at DR21 \& DR21\,(OH) is cold ($\sim$$30$\,K) and concentrated in the ridge.
The clump volume filling is typically a few percent and the density within a clump is of the order $10^6$\,\percc{}.

3. By fitting the dust emission SED, we constrain the H$_2$ column density to 2--5 $10^{22}$\,\percc{} for DR21\,E{}, W{} and toward FIR1.
At DR21/DR21\,(OH) it rises to above $10^{23}$\,\percc{}. The dust temperatures vary between 31--43\,K and are close
to the temperature of the cold gas.

4. Warm gas (80--150\,K) was found  at all five positions.
The warm gas mass is typically a few percent of the total mass but accounts for most of the total CO emission (80\.\% to 94\,\%, mainly in
the mid- and high-$J$ tranitions).
DR21\,(OH), as the coldest source, still shows a significant amount of high-$J$ CO emission.

5. CS 7\TO6  emission can be reproduced if the local density in the envelope region is sufficiently high.
Therefore, high densities are expected at DR21 and DR21\,(OH) supporting again the low volume filling factors.

6. Supported by the model results, we derived the total intensity of the strongest emission lines (\OI{}, CO, \CII{}, and \CI{}).
Line emission provides up to 2\,\% to the total cooling of the atomic and molecular gas and by dust grains.
At the dense cores DR21 and DR21\,(OH), the cooling efficiency is less than 0.5\,\%.
At DR21, \OIw{} and CO are equally strong, whereas
at DR21\,(OH) the CO lines dominate the cooling with more than 75\,\%.
\CIIw{} contributes generally less than 20\,\% and \CIw{} not more than 1.5\,\% to the line cooling.
The \OI{} emission shows a large variation from $<15$\,\% at DR21\,(OH) and FIR1 to $61$\,\% at E.



\begin{acknowledgements}
  The KOSMA 3m submillimeter telescope at the Gornergrat-S\"ud
  is operated by the University of Cologne in collaboration with the
  Radio Astronomy Department of the Argelander-Institute for Astronomy (Bonn), and supported by
  special funding from the Land NRW.  The observatory is administrated
  by the International Foundation Gornergrat \& Jungfraujoch.

\end{acknowledgements}

\bibliographystyle{aa}
\bibliography{aamnem99,aa5855-04}


%

\appendix

\section{Results at individual positions}
\label{rad-model_pos}

Table \ref{tab-carbon-content} lists the derived physical parameters (shell- and model-averaged)
for temperature, H$_2$ number density, volume filling factors,
H$_2$ column density, and molecular gas mass.
The resulting spectral profiles are compared with the observed lines in Fig. \ref{plot-spectra}.
The plots in Figure \ref{fig_cooling} show the fit of the integrated emission for
CO, $^{13}$CO, and C$^{18}$O up to $J=20$ in comparison to the observed lines.
In the following, we discuss details of the individual positions:

\paragraph{DR21\,C{}:}

This position covers the DR21 \HII{}-region. Its surrounding molecular cloud (within a radius of $\sim$0.6\,pc) forms the southern part of the molecular ridge.
Our best-fit model indicates that the envelope at a temperature of 40\,K
comprises roughly 94\,\% of the total mass, leaving 6\,\% for the warm (about 150\,K) and dynamically active gas
($v_\RM{turb}$ between 4.5 and 29$\,\kms$)
subject to shocks and UV-heating.
\citeauthor{jaffe1989} already found in \citeyear{jaffe1989} indications of a warm component with a mass of 55\,\msol{}(D/1.7\,kpc)$^2$, which is fully consistent with our result from Table \ref{tab-carbon-content}.
The by far more massive, colder gas component was only indirectly traced by the CO $J$=7\TO6 line reversal in their study.
In our model, the properties of this gas component in the envelope shell is much better constrained
by the C$^{18}$O and $^{13}$CO $J$=$1\TO0$ transitions which are close to optically thin ($\tau\sim 0.1$ resp. $0.8$).
The turnover at $J_\RM{upper}\sim12$ (cf. Figure~\ref{fig_cooling}) results from the
increasing critical density
leading to sub-thermal excitation of the 
high-$J$ transitions
at the clump density of about $3\cdot 10^6$\,\percc{}.
\citet{boreiko1991} observed CO and $^{13}$CO 9\TO8 at comparable resolution onboard the Kuiper Airborne Observatory (KAO).
We indicate their integrated fluxes in Fig. \ref{fig_cooling} for DR21\,C{}.
There is a good agreement with $^{13}$CO, but CO $J$=$9\TO8$ is weaker than the here presented $J$=$7\TO6$ line.
This may indicate a stronger flattening of the CO rotation curve that
could only be explained by a discontinuous density or temperature structure.

In support of a shock driven excitation, \citet{lane1990} gave an upper limit to CO $J$=22\TO21 of $2\cdot 10^{-4}$\,erg s$^{-1}$ cm$^{-2}$ sr$^{-1}$ in a $44''.8$ beam.
Considering beam dilution, this limit is still high compared to the intensity of CO 17\TO16, the highest rotational line considered here.
Our prediction with the radiative transfer model of $2\cdot 10^{-7}$\,erg s$^{-1}$ cm$^{-2}$ sr$^{-1}$
can explain why this detection attempt failed, albeit
we cannot fully exclude the presence of $>200$\,K warm gas.
\OUT{
 (see below).
}

\paragraph{DR21\,E{}:}

The east side of the DR21 \HII{}-region is not located on the molecular ridge, but enriched with
gas ejected from the rigde.
The lack of cold gas gives raise to an average temperature of 62\,K for a gas mass of $\sim$$240$\,\msol.
From the Spitzer/IRAC image it appears that a fraction of gas in the east is undergoing a blister phase.
Unlike the western outflow side, it is not stopped by a dense cloud and expands unhampered.
We find that 20\,\% of the gas is heated to about 90\,K and accelerated to velocities up to $v=32\,\kms$.
As the radiative transfer code is not designed for outflow geometries,
we applied a turbulent velocity gradient from 32 to $5\,\kms$ to simulate the outflow.
For the envelope gas, the asymmetry of the CO profiles indicates a motion away from the core at $\sim{}0.7\,\kms$.
Because of the steep drop of high-$J$ CO emission, the local clump density was fitted
below $n(\RM{H}_2)=10^5$\,\percc{}.

\paragraph{DR21\,W{}:}

Molecular line emission and radiative transfer models of DR21\,W{} are similar to position DR21\,E{}:
The gas is warmer than in the ridge ($>50$\,K) but not at densities above $10^5$\,\percc{}.
The warm ($\sim 120$\,K) gas contributes with a small fraction of about 3.5\,\% to the total derived mass of $\sim 430$\,\msol{}.
A turbulent velocity gradient between 30 and $11\,\kms$ is responsible for line wing emission.
The CO lines show only weak self-absorption although the opacity is high at the line center.
This is plausible because the cloud surface is illuminated by the cluster at DR21\,C{}.
Hence, we had to implement a positive temperature gradient from 44\,K to 55\,K across the cloud envelope.

\OUT{
Shocks as cause of H$_2$ rotational and ro-vibrational transitions in the DR21 outflow
were discussed by \citet{wright1997} based on ISO/SWS observations.
They find that the molecular gas emission at an East and a West position
($30''$ and $60''$ away from the \HII{}-region) stems from two components at more than T$_1=640$\,K and T$_2=1650$\,K,
probably heated by shock and ultra-violet excitation.
In order to test the mass contribution of hot gas in the modeling, we created a hot gas model with a temperature gradient running from
T$_2$ to T$_1$ and fitted the CO high-$J$ line emission.
With a constant density model at $4\times 10^4$\,\percc{} and a volume filling of 0.8\,\%, we can explain the observed high-$J$ line intensities.
However, at this low column with a total mass of $\sim 2$\,\msol{} we cannot explain the strong $^{13}$CO emission in $J=6\TO5$.
Going to even higher densities would further reduce the mass.
A forthcoming paper will discuss the high temperatures in the context of a PDR model.
}

\paragraph{DR21\,(OH):}

The position of DR21\,(OH) is dominated by cold gas from the ridge at a temperature below 30\,K over a size of up to 0.7\,pc.
For the modeling, we simplified the scenario of several embedded cores as reported by, e.g., \citet[][]{mangum1992} (at possibly different systemic velocities),
to a single clumpy core defined by the warm interface region with an outer radius of 0.2\,pc.
The significant mid- and high-$J$ CO emission is a clear indication for warm and dense gas.
We find only a few percent (2\,\%, or 34\,\msol) of the gas is heated to $\sim$$100$\,K and at densities of $10^6$\,\percc{} or more.
There is no evolved outflow associated with this position, but \citet{garden1986} and \citet{richardson1994} see indications
of a very young outflow associated with high temperatures and shocks.
In our model, half of the warm gas mass is at relative velocities $5$$<$$v_\RM{turb}$$<$$32\,\kms$ creating the line wing emission.
The observed blue-skewed CO line profiles are compatible with radial infall motion of the envelope gas of $1.0-1.5\,\kms$.




\paragraph{DR21\,FIR1:}

The northern FIR-region contains at least three sub-sources.
Besides FIR1, the $80$\arcsec{} beam partly covers the second source FIR2, so we have to consider this as contamination.
The modeled outer radius overlaps slightly with the neighbouring source, but is well within the beam.
Because high-$J$ CO is not detected at DR21\,FIR1, we can constrain the temperature to $\leq 82$\,K.
Given the CO 7\TO6 flux at this position, the lower limit of the temperature in the inner region is T$>50$\,K.
The relatively weak self-absorption of the low-$J$ lines is probably due to the presence of moderately warm gas (up to 40\,K) in the outer surface layers,
supporting the flat density profile used in the model.


%
\section{Correction for Reference-Beam emission}
\label{self-chopping}

\begin{figure}[ht]
\begin{center}
\includegraphics[totalheight=0.47\textwidth,angle=-90]{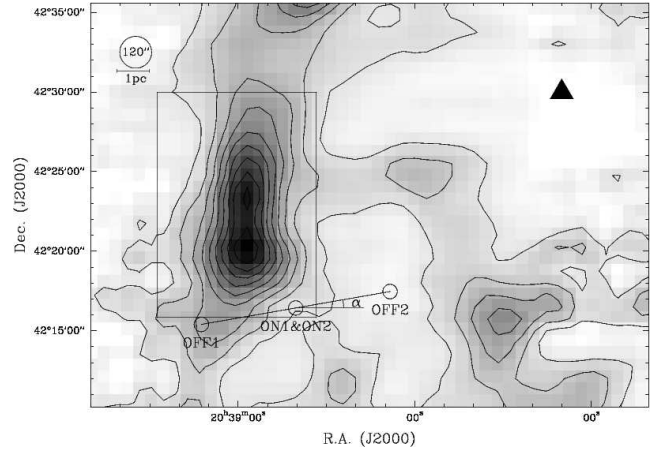}
\end{center}
\caption{On-The-Fly map of the $^{13}$CO 2\TO1 emission \citepalias{schneider2006}.
{\NEW An emission-free position is indicated by the triangle on the right and was used as OFF position
for the OTF observations.}
The box indicates the map size of DBS observations in Fig.~\ref{fig_map_co} and \ref{fig_map_ci}.
The circle in the top left corner indicates the $120''$ beam size.
{\NEW The smaller circles labeled ON1\&ON2, OFF1, and OFF2 show an examplary DBS measurement (cf. Fig. \ref{fig_correction})
with a field rotation angle $\alpha=10^{\circ}$.}
\label{fig_13co21}
}
\end{figure}

\begin{figure}[ht]
\begin{center}
\includegraphics[totalheight=0.4\textwidth,angle=-90]{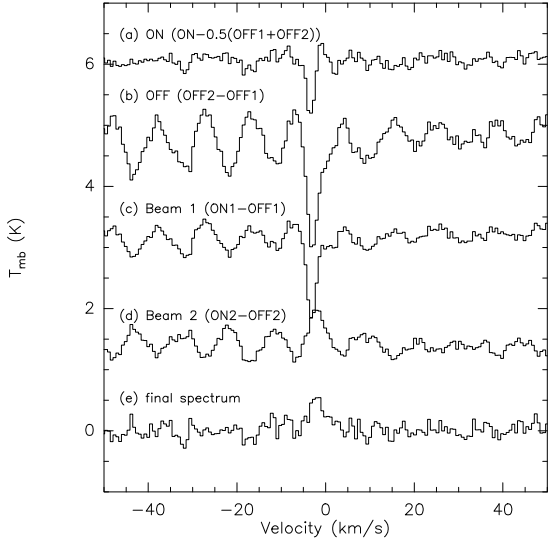}
\end{center}
\caption{This plot demonstrates the method to correct self-chopping in Dual-Beam-Switch observing mode.
The uppermost spectrum {\it (a)} shows a standard DBS spectrum of \CIw{} 1\TO0 with a self-chopped absorption line
(ON$-0.5$(OFF1$+$OFF2)), and {\it (b)} shows the difference signal (OFF2$-$OFF1). 
In the middle, {\it (c)} and {\it (d)}, both beams of the DBS spectrum are displayed (Beam 1 or ON1$-$OFF1, and resp. for Beam 2).
The result {\it (e)} (after removing standing waves) shows a faint emission line.
\label{fig_correction}
}
\end{figure}


Before the implementation of an {\MOD On-The-Fly-mode (OTF)}, Dual-Beam-Switch (DBS) was the only efficient observing mode to use SMART at KOSMA.
{\NEW Therefore, observations in 2003 and 2004 of CO 7\TO6 and \CI{} had to be performed in DBS-mode.
In this mode, }
a chopping secondary mirror is used for fast periodic switching between an ON and an OFF position (Signal- and Reference phase).
{\MOD After some time, usually 20\,sec, determined by the stability of the system, }
the telescope is moved so that the Reference phase is now the ON, and the new OFF
position falls exactly on the opposite side. In the data reduction, each OFF position is weighted equally.
{\NEW Slight differences in the optical path lengths usually lead to strong
standing waves in the spectra when using a simple BS-mode. The period of these waves
corresponds to the optical length to the subreflector. These standing waves are very efficiently suppressed in DBS-mode.}
However, mapping of extended sources is rather difficult if the chopper-throw {\NEW (between $5'-6'$)} is small compared to the size of the source
and the chopper {\NEW movement is fixed to one direction (azimuth for KOSMA)}.
Under those conditions, mapping needs careful planning to prevent self-chopping within the source.
Observing time is thus not only constrained by source visibility but also by sky rotation.
For an hour angle  of $0^\RM{h}$, the secondary throw is only in right ascension, while for other angles there is a component in declination.
The $^{13}$CO 3\TO2 {\MOD Position-Switched OTF-}map in Fig. \ref{fig_13co21} shows that
the DR21 ridge is N-S orientated and field rotation angles {\NEW $\alpha$} of less than $\pm 30^\circ$
are preferred in order to prevent chopping into the ridge of emission with both OFF beams.

To reduce the impact of {\NEW artefacts remaining after the standard DBS reduction pipeline},
all DBS data were corrected in the following way (see Fig.~\ref{fig_correction} for an example):
{\it (i)}
Regardless whether the DBS-spectra (Fig. \ref{fig_correction} (a)) show self-chopping or not,
the emission in the {\it difference} (Fig. \ref{fig_correction} (b)) of the two OFF positions is analysed.
Any significant signal here is considered as pollution. This assumption
holds as long as at least one of the two OFF positions is free of emission and no absorption against the continuum is present.
In the example (Fig. \ref{fig_correction} (c) and (d)) only the OFF-position of Beam 1 (OFF1) is contaminated.
{\it (ii)} If a signal in the difference is detected above the noise level, the algorithm tries to reconstruct the emission signal (Fig. \ref{fig_correction} (d))
by removing Beam 1 from the final spectrum (Fig. \ref{fig_correction} (e)), i.e. we use only ON2-OFF2 here.
This scheme works well for situations where standing waves are negligible.
{\NEW
In cases where standing-wave patterns are persistent, we transform the spectra to fourier-space, mask-out the standing wave frequencies,
and transform the data back to user-space}.
Due to the shorter integration time, this procedure increases the noise level by a factor of $\sqrt2$.
{\NEW In case of DR21, this method successfully recovered the emission because one OFF position was
always nearly free of emission. If, however, the source extends over more than twice the chop-throw
in all directions this method does not work.}

\section{LTE Methods}
\label{methods}

A simple determination of the excitation temperature and the column density of \CIw{} can be done
if we assume the emission is optically thin and the level popolation is given by thermal excitation.
The column density of a material in an excited state $u$ is given by:
\begin{equation}
N_{u}=\frac{8\pi k_\RM{B} \nu^2}{hc^3A_{u}}\int \frac{T_\RM{mb}\RM{d}v}{\kkms}.
\end{equation}
With Einstein A coefficients for atomic carbon
$A_{1}=7.9\times 10^{-8}\RM{s}^{-1}$ and
$A_{2}=2.7\times 10^{-7}\RM{s}^{-1}$
\citep{schroeder1991}
the column densities for each transition are
\begin{equation}\label{cicolumn1}
N_{1}=5.96\,{}10^{15}\int \frac{T_\RM{mb}(\ion{C}{i}~^3\RM{P}_1\TO^3\RM{P}_0)\RM{d}v}{\kkms}~[\RM{cm}^{-2}]
\end{equation}
and
\begin{equation}\label{cicolumn2}
N_{2}=4.72\,{}10^{15}\int \frac{T_\RM{mb}(\ion{C}{i}~^3\RM{P}_2\TO^3\RM{P}_1)\RM{d}v}{\kkms}~[\RM{cm}^{-2}].
\end{equation}

To derive the exitation temperature, we assume LTE
and substitute $N_{1}$ and $N_{2}$ using Boltzmann statistics:
\begin{eqnarray}\label{tempratio}
&&\frac{\RM{N}_u}{\RM{N}_l}=\frac{g_{u}}{g_{l}}e^{-\frac{(E_u-E_l)/k_B}{T_\RM{ex}}}\\
T_\RM{ex}&=&\frac{\frac{h\nu_{2}}{k_B}}{\ln(\frac{N_{1}}{N_{2}}\frac{g_{2}}{g_{1}})}=
\label{ciratio}
\frac{38.8\,\RM{K}}{\ln(\frac{1.26}{R_\RM{\CIw{}}}\frac{5}{3})}=
\frac{38.8\,\RM{K}}{\ln(2.11/R_\RM{\CIw{}})},
\end{eqnarray}
%
%
$R_\RM{\CIw{}}$ is the ratio of integrated intensities (defined in (\ref{cicolumn1}) and (\ref{cicolumn2})) and the
statistical weight is $\RM{g}_J$=$2J+1$.
Although the temperature of the 3-level state of \CI{} is very sensitive to small variations in $R_\RM{\CIw{}}$, the
toal column density depends only weakly on the temperature:
\begin{eqnarray}\label{columnci}
N(\CIw{})&=&\sum_{J=0}^{2}\RM{N}_J=N_1 \frac{Z}{3e^{-\frac{23.6\,\RM{K}}{T_\RM{ex}}}}\\
&=&5.96\,{}10^{15}\frac{1+3e^{\frac{-23.6\,\RM{K}}{T_\RM{ex}}}+5e^{\frac{-62.4\,\RM{K}}{T_\RM{ex}}}}{3e^{\frac{-23.6\,\RM{K}}{T_\RM{ex}}}}\times\nonumber\\
&&\int \frac{T_\RM{mb}(\ion{C}{i}~^3\RM{P}_1\TO^3\RM{P}_0)\RM{d}v}{\kkms}~[\RM{cm}^{-2}]
\end{eqnarray}
where $Z=\sum_{J=0}^{2}\RM{g}_J e^{-\frac{E_J}{k_\RM{B}T_\RM{ex}}}$ is the partition function.

To extimate the excitation temperature from the two transitions $^{13}$CO $J$=3\TO2 and $J$=1\TO0
we used Eq. (\ref{tempratio}) which then translates into
\begin{eqnarray}\label{coratiolow}
T_\RM{ex,low}&=&-\frac{(E_3-E_1)/k_B}{\ln{(\frac{N_3}{N_1}\frac{g_1}{g_3})}}=\frac{26.4\,\RM{K}}{\ln{(9/R_{^{13}\RM{CO,low}})}}.
\end{eqnarray}
The transition $J$=6\TO5 and $J$=3\TO2 gives accordingly
\begin{eqnarray}\label{coratio}
T_\RM{ex,high}&=&\frac{79.4\,\RM{K}}{\ln{(4/R_{^{13}\RM{CO,high}})}}.
\end{eqnarray}
$R_{^{13}\RM{CO,low}}$ (resp. $R_{^{13}\RM{CO,high}}$) is the integrated intensity ratio of
3\TO2/1\TO0 (resp. 6\TO5/3\TO2).


The $^{13}$CO and C$^{18}$O column density can then be approximated by these equations
\begin{eqnarray}\label{column13co}
N(^{13}\RM{CO})&=&1.268\cdot 10^{12} (T_\RM{ex} + 0.88\,\RM{K}) e^{111.1\,\RM{K}/T_\RM{ex}} \times\nonumber\\
&&\int \frac{T_\RM{mb}(^{13}\RM{CO} (6\TO5) ) \RM{d}v}{\kkms}~[\RM{cm}^{-2}],
\end{eqnarray}

\begin{eqnarray}\label{column13colow}
N(^{13}\RM{CO})&=&4.71\cdot 10^{13} (T_\RM{ex} + 0.88\,\RM{K}) e^{5.27\,\RM{K}/T_\RM{ex}} \times\nonumber\\
&&\int \frac{T_\RM{mb}(^{13}\RM{CO} (1\TO0) ) \RM{d}v}{\kkms}~[\RM{cm}^{-2}],
\end{eqnarray}

\begin{eqnarray}\label{columc18o}
N(\RM{C}^{18}\RM{O})&=&4.60\cdot 10^{13} (T_\RM{ex} + 0.88\,\RM{K}) e^{5.27\,\RM{K}/T_\RM{ex}} \times\nonumber\\
&&\int \frac{T_\RM{mb}(\RM{C}^{18}\RM{O} (1\TO0) ) \RM{d}v}{\kkms}~[\RM{cm}^{-2}].
\end{eqnarray}
Using Eq. (\ref{tempratio}), an upper limit to the excitation temperature of \OIw{} can be derived from the
\OI{} 145/63\,\micron{} ratio (fluxes in erg\,s$^{-1}$\,cm$^{-2}$ sr$^{-1}$) by:
\begin{eqnarray}\label{oiratio}
\RM{T}_\RM{ex}&=&\frac{227.7\,\RM{K}}{\ln(2.83/(12.22\cdot R_\RM{\OI{}}))}.
\end{eqnarray}

\OUT{
The column densities of $^{13}$CO and C$^{18}$O were derived from $N(CO)=f(\RM{T}_\RM{ex})\int\RM{T}_\RM{MB}\RM{d}v\,[\kkms]$ with

\begin{equation}
f(\RM{T}_\RM{ex})=\frac{3h}{8\pi^3\mu_\RM{CO}^2}\frac{Z}{J}\frac{\exp{\frac{h\RM{B}_\RM{CO}J(J-1)}{k_B\RM{T}_\RM{ex}}}}{\left(1-\exp{\frac{-h\nu_{J,J-1}}{k_B\RM{T}_\RM{ex}}}\right)}
(J_\nu(\RM{T}_\RM{ex}))^{-1}
\end{equation}
\begin{equation}
Z=\sum_{J=0}^\infty(2J+1)\exp{\frac{-BJ(J+1)hc}{k_B\RM{T}_\RM{ex}}}
\end{equation}

For $Z(^{13}\RM{CO})=k_B/h\RM{B}_{^{13}\RM{CO}}\cdot\RM{T}_\RM{ex}+1/3$

The total column density was then derived by fitting the above formula using $N=X\cdot(\RM{T}_\RM{ex}+hB/3k_B)\exp{\frac{hBJ(J-1)}{k_B\RM{T}_\RM{ex}}}$

}

\end{document}